\documentclass[10pt]{article}

\usepackage{amsmath}
\usepackage{amssymb}
\usepackage{graphicx}
\usepackage{epic,eepic}

\usepackage{color}

\usepackage{amstext,amsthm,amssymb,amsmath}
\usepackage{graphicx}
\usepackage{subfigure}
\usepackage{wrapfig}
\usepackage{fullpage}
\usepackage{color}
\usepackage{multirow}
\usepackage{tabulary}
\usepackage{booktabs}
\usepackage{enumerate}
\usepackage{epstopdf}

\newtheorem{theorem}{Theorem}[section]

\newtheorem{remark}[theorem]{Remark}

\hfuzz=\maxdimen
\graphicspath{ {./pics-2/}} 

\title{Multiscale model reduction for shale gas transport in fractured media}

\author{I. Y. Akkutlu \thanks{Department of Petroleum Engineering, Texas A \& M University, College Station, TX} \and 
Yalchin Efendiev\thanks{Department of Mathematics, Texas A\&M University, College Station, TX.} \and 
Maria Vasilyeva\thanks{Department of Computational Technologies, Institute of Mathematics and Informatics, North-Eastern Federal University, Yakutsk, Republic of Sakha (Yakutia), Russia, 677980 \& Institute for Scientific Computation, Texas A\&M University, College Station, TX 77843-3368}
}

\begin{document}
\maketitle

\begin{abstract}

In this paper, we develop a multiscale model reduction technique that describes shale gas transport
in fractured media. Due to the pore-scale heterogeneities and processes, we use upscaled models
to describe the matrix. We follow our previous work \cite{aes14}, where we derived an upscaled
model in the form of generalized nonlinear diffusion model to describe the effects of kerogen.
To model the interaction between the matrix and the fractures, we use Generalized Multiscale Finite Element
Method \cite{egh12, efendiev2015hierarchical}. 
In this approach, the matrix and the fracture interaction is modeled via local multiscale basis
functions. 
In \cite{efendiev2015hierarchical}, we developed the GMsFEM and applied
for linear flows with horizontal or vertical fracture orientations on a 
Cartesian fine grid.
In this paper, we consider arbitrary fracture orientations and use triangular
fine grid and developed GMsFEM for nonlinear flows. Moreover, we develop
online basis function strategies to adaptively improve the convergence.
The number of multiscale basis functions in each coarse region represents the
degrees of freedom needed to achieve a certain error threshold. Our approach is adaptive in a sense that the multiscale
basis functions can be added in the regions of interest. Numerical results for two-dimensional problem are presented to
demonstrate the efficiency of proposed approach.
\end{abstract}

\section{Introduction}

Shale gas transport is an active area of research due to a growing interest in 
producing natural gas from source rocks. 
The shale systems have added complexities due to the presence 
of organic matter, 
known as kerogen. The kerogen brings in new fluid storage and 
transport qualities to the shale. 
A number of authors, e.g., 
Loucks et al. (2009), Sondergeld et al. (2010), and Ambrose et al. (2012),
 \cite{loucks2009morphology, sondergeld2010micro, ambrose2012shale}, 
have previously discussed the physical properties of the kerogen  
using scanning electron microscopy (SEM) and showed
the co-existence of nanoporous kerogen and microporous conventional 
inorganic rock materials. 

Gas transport in the kerogen typically develops at low Reynolds number and 
relatively high Knudsen number values. Under these conditions, it is expected that 
the transport is not driven by laminar (Darcy) flow dominantly but instead by the pore 
diffusion and other molecular transport mechanisms such as Knudsen diffusion and the 
adsorbed phase (or surface) diffusion. The latter introduces nonlinear processes at the
pore scale that occur in heterogeneous pore geometry.
Some types of upscaled models are needed to represent these complex processes for reservoir simulations.

In large-scale simulations, the complex pore-scale transport needs to be coupled to 
the transport in fractures. This brings an additional difficulty in multiscale simulations.
In particular, the multiscale simulations of the processes describing the interaction
between the fracture and the matrix require reduced-order model approaches that
work for problems without scale separation and high contrast. The objective
of this paper is to discuss the development of such approaches for describing the fracture
and the matrix interaction by taking the upscaled matrix model
following our previous work \cite{aes14}.

In our previous work \cite{aes14}, we proposed a set of macroscopic models that 
take into account the nanoporous nature and nonlinear processes of the shale matrix.  
Our derivation uses multiple scale asymptotic analysis applied to mass balance equations, 
equation of state (for free gas) and isotherm of adsorption. The fine-scale microscopic 
description is largely based on the model formulated by Akkutlu and Fathi (2012), 
\cite{akkutlu2012multiscale}. 
The macroscopic parameters that appear in the equations require solutions of 
cell problem defined in representative volume elements (RVEs). These RVE problems 
take into account fine-scale variations and average their effects on macro scale. 

The multiscale approaches proposed in \cite{aes14} are limited to 
representing the features that have scale separation. To represent the fracture
network and the interaction between the fracture network and the matrix,
we present a multiscale approach following the framework of 
Generalized Multiscale Finite Element Method (GMsFEM), \cite{egh12}. 
The main idea of GMsFEM is to use multiscale basis
functions to extract an essential information in each coarse grid (computational
grid) and develop a reduced-order model.
In \cite{efendiev2015hierarchical}, we have developed the GMsFEM and applied
for linear flows with horizontal or vertical fracture orientations on a 
Cartesian fine grid. In this paper, our contributions are: (1) 
the use of  arbitrary fracture orientations and use triangular
fine grids; (2) the development of  GMsFEM for nonlinear flows;
and (3) the development of
online basis function strategies to adaptively improve the convergence.

To represent the fractures on the fine grid, 
we use Discrete Fracture Model (DFM) \cite{yao2013numerical}.
The fine grid is constructed to resolve
the fractures. For the coarse grid, we choose a rectangular grid.
The GMsFEM framework uses these fine-scale models in computing the snapshot
space and the offline space. The nonlinear models are handled with GMsFEM
by locally updating multiscale basis functions.

The study of flows in fractured media has a long
history. Some modeling techniques on the fine grid include
the Discrete Fracture Model (DFM),
Embedded Fracture Model (EFM) \cite{Lough97, Li08, Lee01}, 
the single-permeability model,
the multiple-permeability models 
(\cite {wu2011multiple,Baca84, Karimi-Fard03, Lee01,Hajibeygi11(2),shu06, reichenberger2006mixed,dietrich2005flow}),
and hierarchical fracture models 
 \cite{Lee01}.
Though these approaches are designed for fine-scale simulations,
a number of these approaches represent the fractures at a macroscopic level.
For example, multiple-permeability models represent the network of connected
fractures macroscopically by introducing several permeabilities in each block.
The EFM (\cite{Lough97, Li08, Lee01}) models the interaction of fractures with
the fine-grid blocks separately for each block.
The main idea of hierarchical fracture modeling 
presented in \cite{Lee01}
is to
homogenize small-length fractures
(with the length smaller than the coarse block),
while to represent the large-length fractures.
Some of these approaches can be generalized by incorporating the interaction
of fractures and permeability heterogeneities locally,
which can lead to efficient upscaling techniques, \cite{Durlofsky91, gong08}.

In recent papers \cite{hajibeygi2011loosely},
several multiscale approaches
are proposed for representing the fracture effects. These approaches share
common concepts with the methods that we discuss here in a sense that they
add new degrees of freedom to represent the fractures on a coarse grid.
The main difference is that our approaches use local spectral problems
accompanied by adaptivity to detect the regions,
 where to add new basis functions.
In this regard, the procedure of finding multiscale basis functions
and the enrichment procedure is different from existing techniques.


The proposed method constructs multiscale basis functions
by appropriately selecting
local snapshot space and the local spectral problems for the
underlying nonlinear problem.
The local
spectral problems allow us to adaptively 
enrich in the regions with larger errors. In the paper, we discuss
adaptivity issues and how to add multiscale basis functions in some 
selected regions.
To reduce the computational cost associated with constructing the snapshot
space, we follow \cite{randomized2014} and use randomized boundary conditions.
One of other novel components of the paper is the use of online basis functions
(see \cite{chung2015residual} 
for online basis functions for steady state problems)
for the time-dependent nonlinear problems. The online basis functions
are constructed during the simulation using the residual and they can
reduce the error significantly. These basis functions are used if the offline
basis functions can not reduce the error below a desired threshold.

We present numerical results for some representative examples. In these examples,
we use nonlinear matrix and fracture models. Our numerical results show that
the coarse-scale models with a fewer degrees of freedom can be used to get an accurate
approximation of the fine-scale solution. 
In particular, only 10 \% degrees of the freedom are needed to obtain an accurate
representation of the fine-scale solution. We also add a geomechanical 
contribution to the permeability term, where the permeability depends on the pressure. Furthermore, we demonstrate the use of online basis functions
and how they can reduce the error.

The paper is organized as follows. In the next section, we present a model problem. In Section
3, we discuss the fine-scale model. Section 4 is devoted to the development of GMsFEM, in particular, the offline spaces. In this section, we 
present numerical results for offline basis functions.
In Section 5, we discuss randomized snapshot spaces and show that their
use can give similar accuracy while for less computational cost.
In Section 6, we develop online basis functions and present numerical
results.


\section{Model Problem}
\label{prelim}

In this paper, we will study nonlinear
gas transport in fractured media motivated by several
applications including shale gas.
We are interested in the shale gas transport 
 described in \cite{akkutlu2012multiscale}. Similar equations
arise in other models, where one considers a free gas in the tight
reservoirs. We will consider  general equations
\begin{equation}
\label{eq:model1}
a_{m}(c) {\partial c \over \partial t} = \text{div}(b_{m}(c,x) \nabla c),
\end{equation}
where $c$ is the amount of free gas and $a(c)$ and $b(c)$ contain terms related to storage and adsorption coefficients. 
In \cite{akkutlu2012multiscale}, the authors consider the nonlinear terms have the forms
\[
a_{m}(c) 		= \phi 	+ (1-\phi) \gamma \frac{\partial F}{\partial c}, \quad 
b_{m}(c,x) 	= \phi D + (1-\phi) \gamma D_s \frac{\partial F}{\partial c} + \phi \frac{\kappa}{\mu} R T c,
\]
where $\gamma$ is a parameter, which is unity in kerogen and is equal to $V_{grain,k}/V_{grain}$ in the inorganic material ($V_{grain,k}$ is grain volume and $V_{grain}$ is kerogen grain volume). 
Diffusivity $D$ and porosity $\phi$ are defined for the free fluid in the inorganic matrix and in the kerogen as follows
\[
D =
\begin{cases}
D_k & \text{ in kerogen}  \\ 
D_i & \text{ in inorganic matrix}
\end{cases}, \quad
\phi =
\begin{cases}
\phi_k & \text{ in kerogen}  \\
\phi_i & \text{ in inorganic matrix} 
\end{cases}.
\]

For the free gas we have ideal gas assumption. 
The Darcy law of free gas flow in inorganic matrix is used with permeability $\kappa$ and gas viscosity $\mu$. For the sorbed gas we can use Langmuir or Henry`s  isotherms $F = F(c)$. 
In \cite{wuli2013generalized, yao2013numerical}, the authors discuss a general framework, where
the equations also include nonlinear diffusivity due to
adsorbed gas  in a shale formation. In \cite{lee2012modeling}, the 
nonlinear terms appear due to barotropic effects.

The nonlinear flows also contain components that are 
due to diffusion in the fractures.
One needs additional
equations for modeling  fractures. 
The fractures have high conductivity.
We will use a general equation of the form
\begin{equation}
\label{eq:model2}
a_{f}(c) {\partial c \over \partial t} = \text{div}(b_{f}(c,x) \nabla c)
\end{equation}
to describe the flow within fractures.
In \cite{akkutlu2012multiscale}, the authors use
\[
a_{f}(c)  = \phi_f, \quad 
b_{f}(c,x) = \frac{\kappa_f}{\mu} R T c,
\]
where $\phi_f$ and $\kappa_f$ are the fracture porosity and permeability. 
These problems are solved on a fine grid using DFM as will be described
in Section \ref{sec:fine_grid_disc}. 

In many shale gas examples, the matrix heterogeneities can  be upscaled
and the resulting upscaled equation has the form (\ref{eq:model1}).
However, the interaction between matrix and fractures require some
type of multiscale modeling approach, where the effects of the fractures
need to be captured more accurately. Approaches, 
such as multicontinuum \cite{wuli2013generalized}
are often used, but these approaches use idealized assumptions on fracture
distributions. In this paper, we will use multiscale basis functions 
to represent fracture effects. In our previous work \cite{efendiev2015hierarchical}, we have considered
similar approaches for single-phase flow when fractures (which could be horizontal or vertical) are aligned with 
Cartesian grid. In this paper, we consider arbitrary fracture distribution
in the context of nonlinear flow equations.

The overall model equations will be solved on a coarse grid. Next,
we introduce the concepts of fine and coarse grids.
Let $\mathcal{T}^H$ be a usual conforming partition of the computational domain
$\Omega$ into finite elements (triangles, quadrilaterals, tetrahedra, etc.). 
We refer to this
partition as the coarse grid and assume that each coarse element is partitioned into
 a connected union of fine grid blocks. The fine grid partition will be denoted by
$\mathcal{T}^h$, and is by definition a refinement of the coarse grid $\mathcal{T}^H$.
We use $\{x_i\}_{i=1}^{N}$ (where $N$ denotes the number of coarse nodes) to denote the vertices of
the coarse mesh $\mathcal{T}^H$ and define the neighborhood of the node $x_i$ by
\begin{equation} \label{neighborhood}
\omega_i=\bigcup\{ K_j\in\mathcal{T}^H; ~~~ x_i\in \overline{K}_j\}.
\end{equation}
\begin{figure}[htb]
  \centering
  \includegraphics[width=0.6 \textwidth]{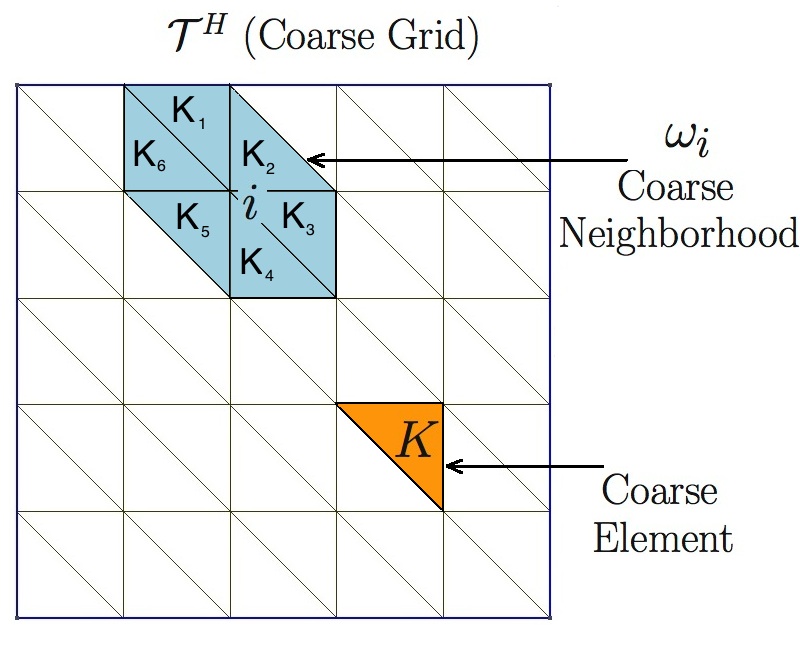}
  \caption{Illustration of a coarse neighborhood and coarse element}
  \label{schematic}
\end{figure}
See Figure~\ref{schematic} for an illustration of neighborhoods and elements subordinated to the coarse discretization. We emphasize the use of $\omega_i$ to denote a coarse neighborhood, and $K$ to denote a coarse element throughout the paper.

\section{Fine-scale discretization}
\label{sec:fine_grid_disc}
To discretize the system on fine grid, we will use finite element method
and use DFM for fractures. 
To solve Problem \eqref{eq:model1} using finite element method (FEM), we need a fine grid discretization to capture the fractures. These computations can be expensive. Here, we apply the discrete fracture network (DFM) model for modeling flows in fractures \cite{Noorishad82}.

In the discrete-fracture model, the aperture of the fracture appears as a factor in front of the one dimensional integral for the consistency of the integral form. This is the main idea of the discrete-fracture model, which can be applied in any complex configuration for fractured porous media.

\begin{figure}[htb]
  \centering
  \includegraphics[width=0.45 \textwidth]{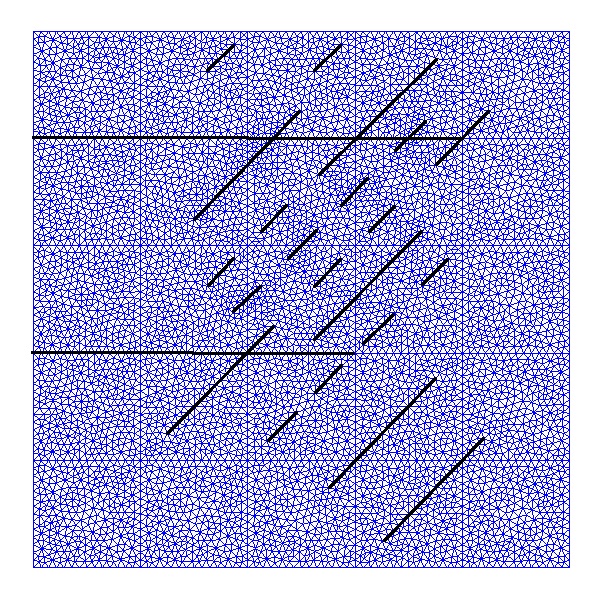}
  \caption{Fine grid with fractures}
  \label{fig:frame_int}
\end{figure}

To demonstrate it, we consider the two-dimensional problem of Equation \eqref{eq:model2}. We simplify the fractures as the lines with small aperture. Thus, one-dimensional element is needed to describe fractures in the discrete-fracture model. The system of equations \eqref{eq:model1} will be discretized in a two-dimensional form for the matrix and in one-dimensional form for the fractures. The whole domain $\Omega$ can be represented by
\begin{equation}
\Omega = \Omega_m \oplus_i \, d_i \Omega_{f,i},
\end{equation}
where $m$ and $f$ represent the matrix and the fracture of the permeability field $\kappa$, respectively. Here, $d_i$ is the aperture of the $i$ th fracture and $i$ is the index of the fractures. Note that $\Omega_m$ is a two-dimensional domain and $\Omega_{f,i}$ is a one-dimensional domain (Figure \ref{fig:frame_int}).
Then Equations \eqref{eq:model1} and \eqref{eq:model2}  and can be written as follows (for any test function $v$):
%
\begin{equation} 
\label{eq:discr1}
\begin{split}
m( \frac{\partial c}{\partial t}, v) +
a(c, v) &=
\int_{\Omega_m}   a_{m}(c) {\partial c \over \partial t} v \, dx +
d_i \sum\limits_i 
\int_{\Omega_{f,i}} a_{f}(c) {\partial c \over \partial t} v \, dx +\\
&+\int_{\Omega_m}   b_{m}(c,x)\nabla c \cdot \nabla v \, dx +
d_i \sum\limits_i 
\int_{\Omega_{ f,i}} b_{m}(c,x)\nabla c\cdot \nabla v \, dx = 0.
\end{split}
\end{equation}

To solve (\ref{eq:discr1}), we will first linearize the system. We will use
the following linearization
\begin{equation}
\label{eq:discr2}
\begin{split}
m( {c^{n+1} - c^{n} \over \tau}, v) +
a(c^{n+1}, v) &=
\int_{\Omega_m}   a_{m}(c^{n}) { c^{n+1} - c^{n} \over \tau} v \, dx +
d_i \sum\limits_i 
\int_{\Omega_{f,i}} a_{f}(c^{n}) { c^{n+1} - c^{n} \over \tau} v \, dx + \\
&+\int_{\Omega_m}   b_{m}(c^{n},x)\nabla c^{n+1} \cdot \nabla v \, dx +
d_i \sum\limits_i 
\int_{\Omega_{ f,i}} b_{f}(c^{n},x)\nabla c^{n+1}\cdot \nabla v \, dx = 0.
\end{split}
\end{equation}
The standard fully-implicit finite difference scheme is used for the approximation with time step size $\tau$ and superscripts $n$, $n+1$ denote previous and current time levels. This is a first-order in time and 
unconditionally stable linearization.

For standard Galerkin finite element method, we write the solution as $c = \sum_{i = 1}^{N_f} c_i \phi_i$, where $\phi_i$ are the standard linear element basis functions defined on $\mathcal{T}^h$ and $N_f$ denotes the number of the nodes on the fine grid. The equation (\ref{eq:discr2}) can be presented in matrix form:
\begin{equation}
\label{eq:discr3}
M^n \frac{c^{n+1} - c^n}{\tau} + A^n c^{n+1} = 0,
\end{equation}
where $M$ is the mass matrix given by 
\[
M^n = [m_{ij}] = 
\int_{\Omega_m}   a_{m}(c^{n})  \phi_i \phi_j \, dx + 
d_i \sum\limits_i 
\int_{\Omega_{f,i}} a_{f}(c^{n})  \phi_i \phi_j  \, dx,
\]
and $A$ is the stiffness matrix given by
\[
A^n = [a_{ij}] = 
\int_{\Omega_m}   b_{m}(c^{n},x) \nabla \phi_i \cdot \nabla \phi_j \, dx +
d_i \sum\limits_i 
\int_{\Omega_{ f,i}} b_{f}(c^{n},x) \nabla \phi_i \cdot \nabla \phi_j \, dx.
\]
Hence at each time step we have the following linear problem
\begin{equation}
\label{eq:discr4}
Q^n c^{n+1} = M^n c^{n},
\end{equation}
where $Q^n = \left( M^n + \tau A^n \right)$.
This fine scale discretization yields large matrices of size $N_f \times N_f$.

\section{Coarse-grid discretization using GMsFEM. Offline spaces. }


We use multiscale basis functions to represent the solution space. 
We will consider the continuous Galerkin (CG) formulation and signify $\omega_i$ as the support of basis functions.
We denote the basis functions by $\psi_k^{\omega_i}$, which is supported in $\omega_i$,
and the index $k$ represents the numbering of these basis functions.
In turn, the CG solution will be sought as 
\[
c_{\text{ms}}(x, t)=\sum_{i,k} c_{k}^i(t) \psi_{k}^{\omega_i}(x).
\]
Once the basis functions are identified, the CG global coupling is given through the variational form
\begin{equation}
\label{eq:globalG} 
m( \frac{\partial c}{\partial t}, v) +
a(c_{\text{ms}},v) = 0, \quad \text{for all} \, \, v\in
V_{\text{off}},
\end{equation}
where  $V_{\text{off}}$ is used to denote the space spanned by those basis functions and 
\[
 m(c, v) = \int_{\Omega_m}   a_{m} c \, v \, dx +
d_i \sum\limits_i 
\int_{\Omega_{f,i}} a_{f} c \, v \, dx, 
\]
\[
a(c, v) = \int_{\Omega_m}   b_{m}\nabla c \cdot \nabla v \, dx +
d_i \sum\limits_i 
\int_{\Omega_{ f,i}} b_{f}\nabla c \cdot \nabla v \, dx.
\]
%

Let $V$ be the conforming finite element space with respect to the fine-scale partition $\mathcal{T}^h$.
We assume $c\in V$ is the fine-scale solution satisfying
\begin{align}
\label{eqn:fine-scale prb}
m( \frac{\partial c}{\partial t}, v) +
a(c, v) = 0, \quad v\in V.
\end{align}


Next, we describe GMsFEM. GMsFEM consists of offline and online stage. In the offline stage we construct multiscale basis functions and after that in the online stage, we solve our problem for any input parameters, such as right hand sides
or boundary conditions.

Offline computations:
\begin{itemize}
\item[] \textit{Step 1.} Coarse grid generation.
\item[] \textit{Step 2.} Construction of the snapshot space that will be used to compute an offline space.
\item[] \textit{Step 3.} Construction of a ``small'' dimensional offline space by performing dimension reduction in the space of local snapshots.
\end{itemize}

Given the computational domain, a coarse grid can be constructed
and local problems are solved on coarse neighborhoods to obtain 
the snapshot spaces.
Then, smaller dimensional offline spaces are obtained from the 
snapshot spaces by dimension reduction
via some spectral problems \cite{egh12, EGG_MultiscaleMOR, ehg04, eglp13oversampling, randomized2014}. 
After that we can solve our problem in the constructed offline space.
Moreover, we will construct online basis functions that are problem dependent and are computed locally based on some local residuals \cite{Chung_adaptive14, chung2015residual}.

We now present the construction
of the offline basis functions
and the corresponding spectral problems
for obtaining a space reduction.
In the offline computation, we first construct a snapshot space $V_{\text{snap}}^{\omega}$.
The snapshot space can be the space of all fine-scale basis functions
or the solutions of some local problems with various choices of boundary conditions.
For example, we can use the following $\kappa$-harmonic extensions to form a snapshot space.
 For each fine-grid function, $\delta_j^h(x)$,
which is defined by
$\delta_j^h(x)=\delta_{j,k},\,\forall j,k\in \textsl{J}_{h}(\omega_i)$, where $\textsl{J}_{h}(\omega_i)$ denotes the fine-grid boundary node on $\partial\omega_i$. For simplicity, we omit the index $i$.
Given a fine-scale piecewise linear function defined on 
$\partial\omega$ (here $\omega$ is a generic coarse element), we define $\psi_{j}^{\omega, \text{snap}}$ by following variational problem
\begin{equation} 
\label{harmonic_ex}
a(\psi_{j}^{\omega, \text{snap}} , v) =
\int_{\omega_{m}}   b_{m}\nabla \psi_{j}^{\omega, \text{snap}} \cdot \nabla v \, dx + d_j \sum\limits_j 
\int_{\omega_{f,j}} b_{f}\nabla \psi_{j}^{\omega, \text{snap}} \cdot \nabla v \, dx
=0 \quad \text{in } \, \omega,
\end{equation}
and $\psi_{j}^{\omega, \text{snap}}=\delta_j^h(x)$ on $\partial\omega$,  $\omega = \omega_{m} \oplus_j \, d_j \omega_{f,j}$.

For brevity of notation, we now omit the superscript $\omega$, yet it is assumed throughout this section that the offline space computations are localized to respective coarse subdomains.
Let $l_i$ be the number of functions in the snapshot space in the region $\omega$, and
$$
V_{\text{snap}} = \text{span}\{ \psi_{j}^{ \text{snap}}:~~~ 1\leq j \leq l_i \},
$$
for each coarse subdomain $\omega$.

Denote
$$
R_{\text{snap}} = \left[ \psi_{1}^{\text{snap}}, \ldots, \psi_{l_i}^{\text{snap}} \right].
$$

In order to construct the offline space $V_{\text{off}}^\omega$, we perform a dimension reduction of the snapshot space using an auxiliary spectral decomposition. The analysis in \cite{egw10} motivates the following eigenvalue problem in the space of snapshots:
\begin{equation} 
\label{offeig}
A^{\text{off}} \Psi_k^{\text{off}} = \lambda_k^{\text{off}} S^{\text{off}} \Psi_k^{\text{off}},
\end{equation}
where
\[
A^{\text{off}} = [a_{mn}^{\text{off}}] = 
\int_{\omega_m}   b_{m} \nabla \psi_m^{\text{snap}} \cdot \nabla \psi_n^{\text{snap}} \, dx +
d_j \sum\limits_j
\int_{\omega_{ f,j}} b_{f} \nabla \psi_m^{\text{snap}} \cdot \nabla \psi_n^{\text{snap}} \, dx
= R_{\text{snap}}^T A R_{\text{snap}},
\] \[
S^{\text{off}} = [s_{mn}^{\text{off}}] = 
\int_{\omega_m}   b_{m} \psi_m^{\text{snap}} \psi_n^{\text{snap}} \, dx +
d_j \sum\limits_j
\int_{\omega_{ f,j}} b_{f}  \psi_m^{\text{snap}} \psi_n^{\text{snap}} \, dx
= R_{\text{snap}}^T S R_{\text{snap}},
\]
where $A$ and $S$ denote analogous fine scale matrices as defined by
\[
A_{ij} = 
\int_{D_m}   b_{m}(c^{n},x) \nabla \phi_i \cdot \nabla \phi_j \, dx + d_i \sum\limits_i 
\int_{D_{ f,i}} b_{f}(c^{n},x) \nabla \phi_i \cdot \nabla \phi_j \, dx, 
\] \[
S_{ij} = \int_{D_m}   b_{m}(c^{n})  \phi_i \phi_j \, dx + 
d_i \sum\limits_i 
\int_{D_{f,i}} b_{f}(c^{n})  \phi_i \phi_j  \, dx,
\]
where $\phi_i$ is the fine-scale basis function.
To generate the offline space, we then choose the smallest $M^{\omega}_{\text{off}}$ eigenvalues from Eq.~\eqref{offeig} and form the corresponding eigenvectors in the space of snapshots by setting
$\psi_k^{\text{off}} = \sum_{j=1}^{l_i} \Psi_{kj}^{\text{off}} \psi_j^{\text{snap}}$ (for $k=1,\ldots, M^{\omega}_{\text{off}}$), where $\Psi_{kj}^{\text{off}}$ are the coordinates of the vector $\Psi_{k}^{\text{off}}$.

Next, we  create an appropriate solution space and variational formulation that for a continuous Galerkin approximation. We begin with an initial coarse space $V^{\text{init}}_0 = \text{span}\{ \chi_i \}_{i=1}^{N}$. Recall that  $N$ denotes the number of coarse neighborhoods. Here, $\chi_i$ are the standard multiscale partition of unity functions defined by
\begin{eqnarray} 
\label{pou}
a(\chi_i , v) =
\int_{\omega_{m}}   b_{m}\nabla \chi_i \cdot \nabla v \, dx + d_j \sum\limits_j 
\int_{\omega_{f,j}} b_{f}\nabla \chi_i \cdot \nabla v \, dx = 0 \quad K \in \omega \\
\chi_i = g_i \quad \text{on} \, \, \, \partial K, \nonumber
\end{eqnarray}
for all $K \in \omega$, where $g_i$ is a continuous function on $\partial K$ and is linear on each edge of $\partial K$.

We then multiply the partition of unity functions by the eigenfunctions in the offline space $V_{\text{off}}^{\omega_i}$ to construct the resulting basis functions
\begin{equation} \label{cgbasis}
\psi_{i,k} = \chi_i \psi_k^{\omega_i, \text{off}} \quad \text{for} \, \, \,
1 \leq i \leq N \, \, \,  \text{and} \, \, \, 1 \leq k \leq M_{\text{off}}^{\omega_i},
\end{equation}
where $M_{\text{off}}^{\omega_i}$ denotes the number of offline eigenvectors that are chosen for each coarse node $i$. We note that the construction in Eq.~\eqref{cgbasis} yields  continuous basis functions due to the multiplication of offline eigenvectors with the initial (continuous) partition of unity. Next, we define the continuous Galerkin spectral multiscale space as
\begin{equation} \label{cgspace}
V_{\text{off}}  = \text{span} \{ \psi_{i,k} : \,  \, 1 \leq i \leq N \, \, \,  \text{and} \, \, \, 1 \leq k \leq M_{\text{off}}^{\omega_i}  \}.
\end{equation}
Using a single index notation, we may write $V_{\text{off}} = \text{span} \{ \psi_{i} \}_{i=1}^{N_c}$, where $N_c =\sum_{i=1}^{N}M_{\text{off}}^{\omega_{i}}$
denotes the total number of basis functions in the space $V_{\text{off}}$. 
We also construct an operator matrix 
\[
R_0^T = \left[ \psi_1 , \ldots, \psi_{N_c} \right],
\] 
where $\psi_i$ are used to denote the nodal values of each basis function defined on the fine grid.

We seek $c_{\text{ms}}(x) = \sum_i c_i \psi_i(x) \in V_{\text{off}}$ such that
\begin{equation} \label{cgvarform}
m( \frac{\partial c_{\text{ms}}}{\partial t}, v) +
a(c_{\text{ms}}, v) = 0 \quad \text{for all} \,\,\, v \in V_{\text{off}}.
\end{equation}

 We note that variational form in \eqref{cgvarform} yields the following linear algebraic system
\begin{equation}
Q^n_0 c_0^{n+1} = M^n_0 c_0^{n},
\end{equation}
where $c_0$ denotes the nodal values of the discrete CG solution, and $Q^n_0 = R_0 Q^n R_0^T$ and  $M_0^n = R_0 M^n$. We also note that the operator matrix may be analogously used in order to project coarse scale solutions onto the fine grid $c^{n+1} =  R_0^T c_0^{n+1}$. In our simulations presented next, we do not update basis  functions. We discuss basis function update in Section \ref{sec:online}.

\subsection{Numerical result}
\label{sec:numerical}

We present numerical results for the coarse-scale solution using offline basis functions. The basis functions of the offline space are constructed following the procedure described above.  Note that, the basis functions are constructed only once at initial time and used for generating 
the stiffness matrix and the right hand side.

We consider the solution of problem with constant  and nonlinear matrix-fracture coefficients in \eqref{eq:discr2}. As constant coefficients
(see previous section) representing matrix and fracture properties,
 we use following   
\[
a_m = 0.8, \quad b_m = 1.3 \cdot 10^{-7} 
\quad \text{and} \quad 
a_f = 0.001, \quad b_f = 1.0.
\]
For nonlinear matrix-fracture coefficients, we use
\begin{equation}
\label{coeff-ab-m}
a_{{m}}(c) 		= \phi 	+ (1-\phi) \frac{\partial F}{\partial c}, \quad 
b_{m}(c,x) 	= \phi D + (1-\phi) D_s \frac{\partial F}{\partial c} + \phi \frac{\kappa}{\mu} R T c,
\end{equation}
and
\begin{equation}
\label{coeff-ab-f}
a_{f}(c)  = \phi_f, \quad 
b_{f}(c,x) = \frac{\kappa_f}{\mu} R T c,
\end{equation}
where $D_k = 10^{-7} [m^2/s]$, $D_i = 10^{-8} [m^2/s]$, $\phi = 0.04$, $T = 413 [K]$, $\mu = 2 \cdot 10^{-5} [kg/(m \, s)]$ and for fractures $k_f = 10^{-12} [m^2]$, $\phi_f = 0.001$. 

As for permeability $\kappa$  in \eqref{coeff-ab-m}, we use constant $\kappa= \kappa_0$ and stress-dependent model $\kappa=\kappa_m$ (see \cite{gangi1978variation,wasaki2014permeability}) with
\[
\kappa_m = \kappa_0 \left( 1 - \left( \frac{p_c - \alpha p}{p_1} \right)^\mathcal{M} \right)^3,
\]
where $\kappa_0 = 10^{-18} [m^2]$, $p = RT c$, $p_c = 10^9 [Pa]$, $p_1 = 1.8 \cdot 10^9 [Pa]$, $\alpha = 0.5$ and $\mathcal{M} = 0.5$.
For the sorbed gas, we use Langmuir model 
\[
F(c)= c_{\mu s} \frac{s}{(1 + s c)^2},
\] 
where $s = 0.26 \cdot 10^{-3}$ and $c_{\mu s} = 0.25 \cdot 10^{-5}  [mol/m^3]$.

\begin{figure}[htb]
  \centering
  \includegraphics[width=0.75 \textwidth]{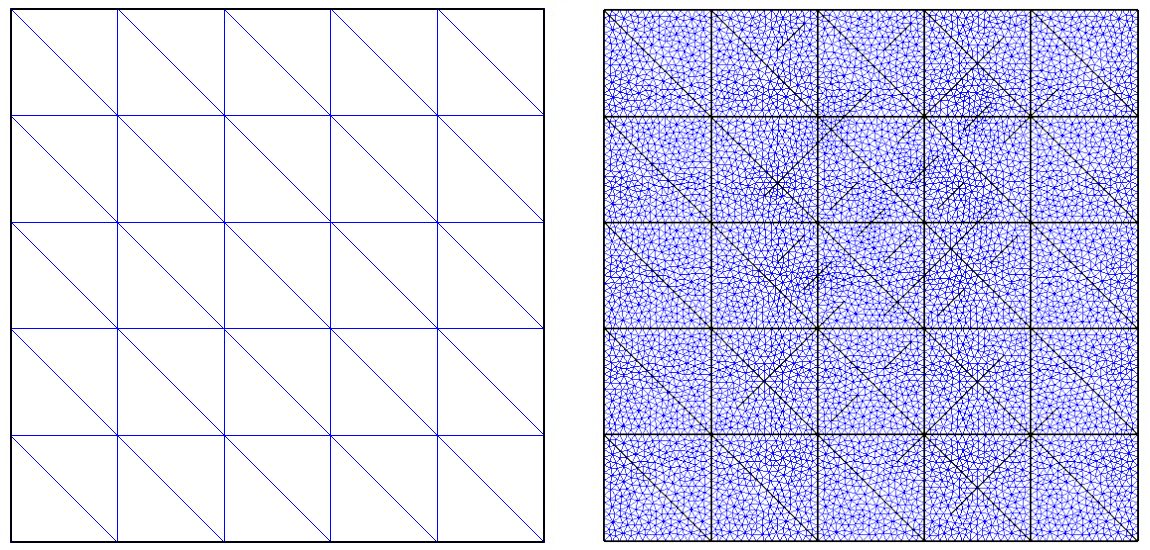}
  \caption{Coarse and fine grids. Coarse grid contains 50 cells, 85 facets and 36 vertices. Fine grid contains 7580 cells, 11470 facets and 3891 vertices. }
  \label{fig:meshes5}
\end{figure}

\begin{figure}[htb]
  \centering
  \includegraphics[width=0.75 \textwidth]{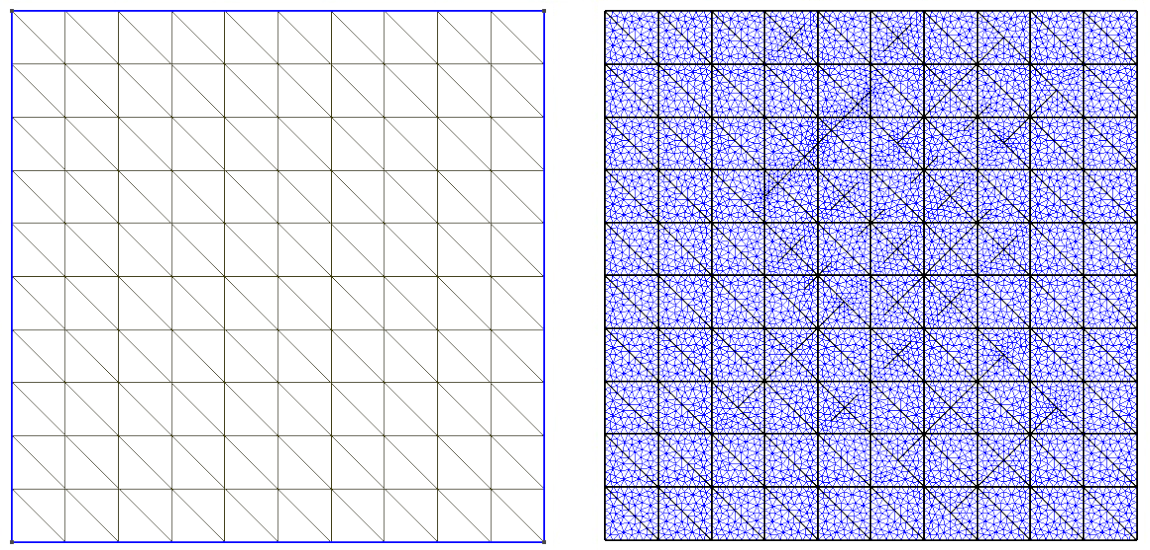}
  \caption{Coarse and fine grids. Coarse grid contains 200 cells, 320 facets and 121 vertices. Fine grid contains 13036 cells, 19694 facets and 6659 vertices. }
  \label{fig:meshes10}
\end{figure}

The equation is solved with Dirichlet boundary condition $c(x, t) = 5000$ on the left boundary and Neumann boundary conditions $\frac{\partial c(x, t)}{\partial n} = 0$ on other boundaries.
The domain $\Omega$ has a length of 60 meters in both directions. We calculate concentration for $t_{max} = 5$ years with the time step $\tau  = 10$ days.  As for initial condition, we use  $c(x, t=0) = 10000 [mol/m^3]$. For the numerical solution, we construct structured two coarse grids with 36 nodes (Figure \ref{fig:meshes5}) and with 121 nodes (Figure \ref{fig:meshes10}). As for fine grids, we use unstructured grids, which resolves the existing fractures.

\begin{figure}[htb]
  \centering
  \includegraphics[width=0.85 \textwidth]{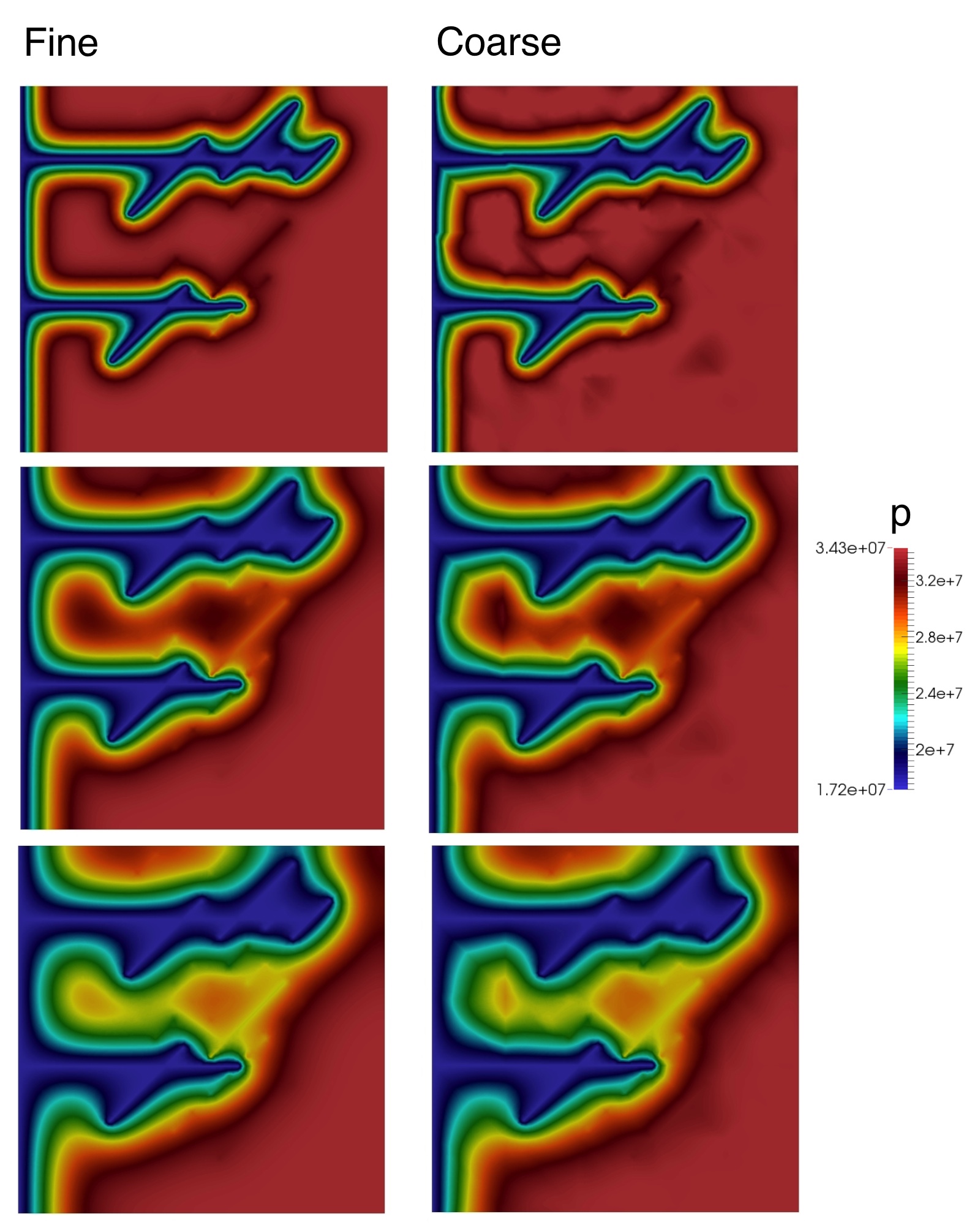}
  \caption{Solution with constant matrix-fracture coefficients on coarse (top) and on fine (bottom) grids for t=1, 3 and 5 year (from top to bottom)}
  \label{fig:solution-lin}
\end{figure}

\begin{figure}[htb]
  \centering
  \includegraphics[width=0.85 \textwidth]{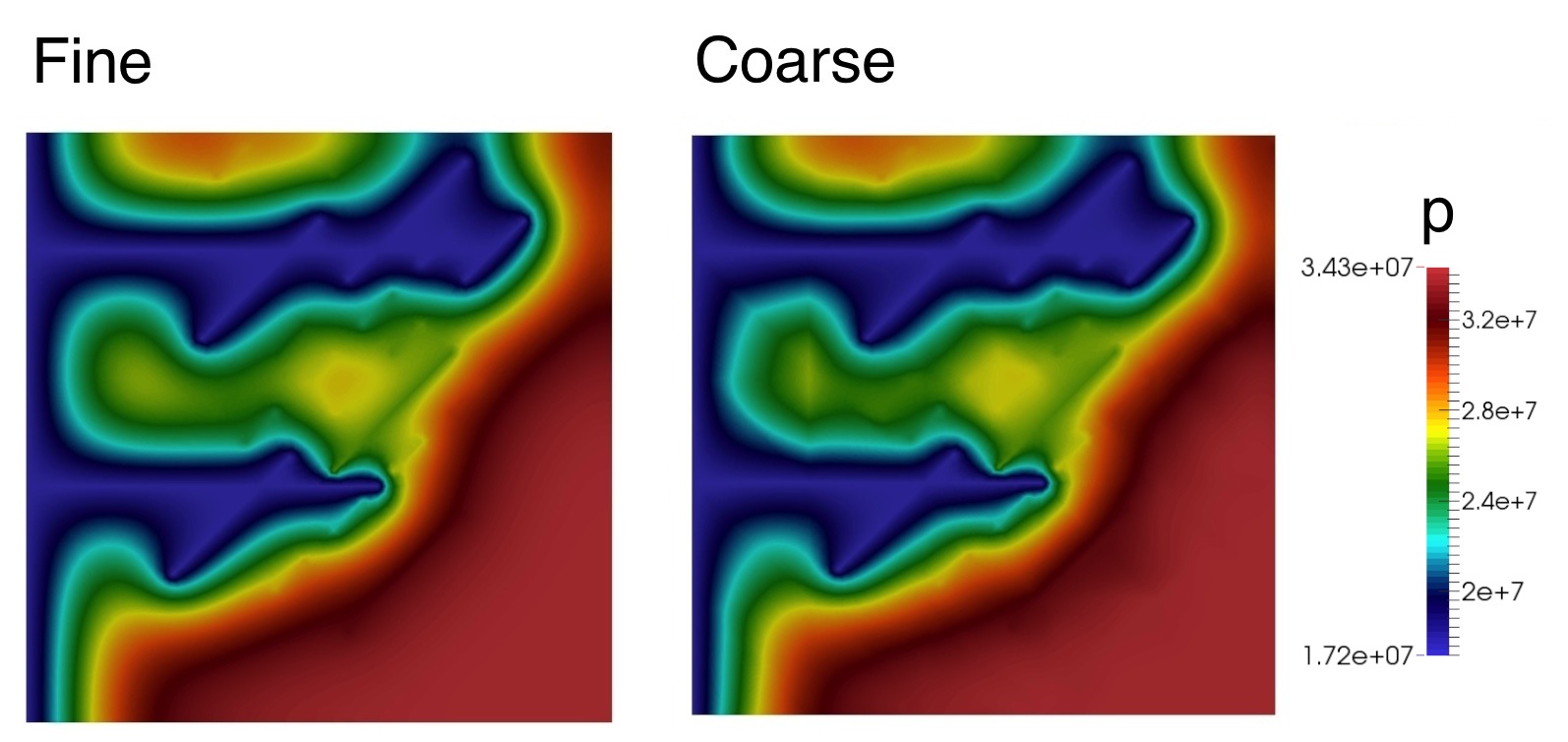}
  \caption{Coarse-scale (top) and fine-scale (bottom) solutions for t=5 year for the case of nonlinear permeability with $\kappa = \kappa_0$}
  \label{fig:solution-k0}
\end{figure}

\begin{figure}[htb]
  \centering
  \includegraphics[width=0.85 \textwidth]{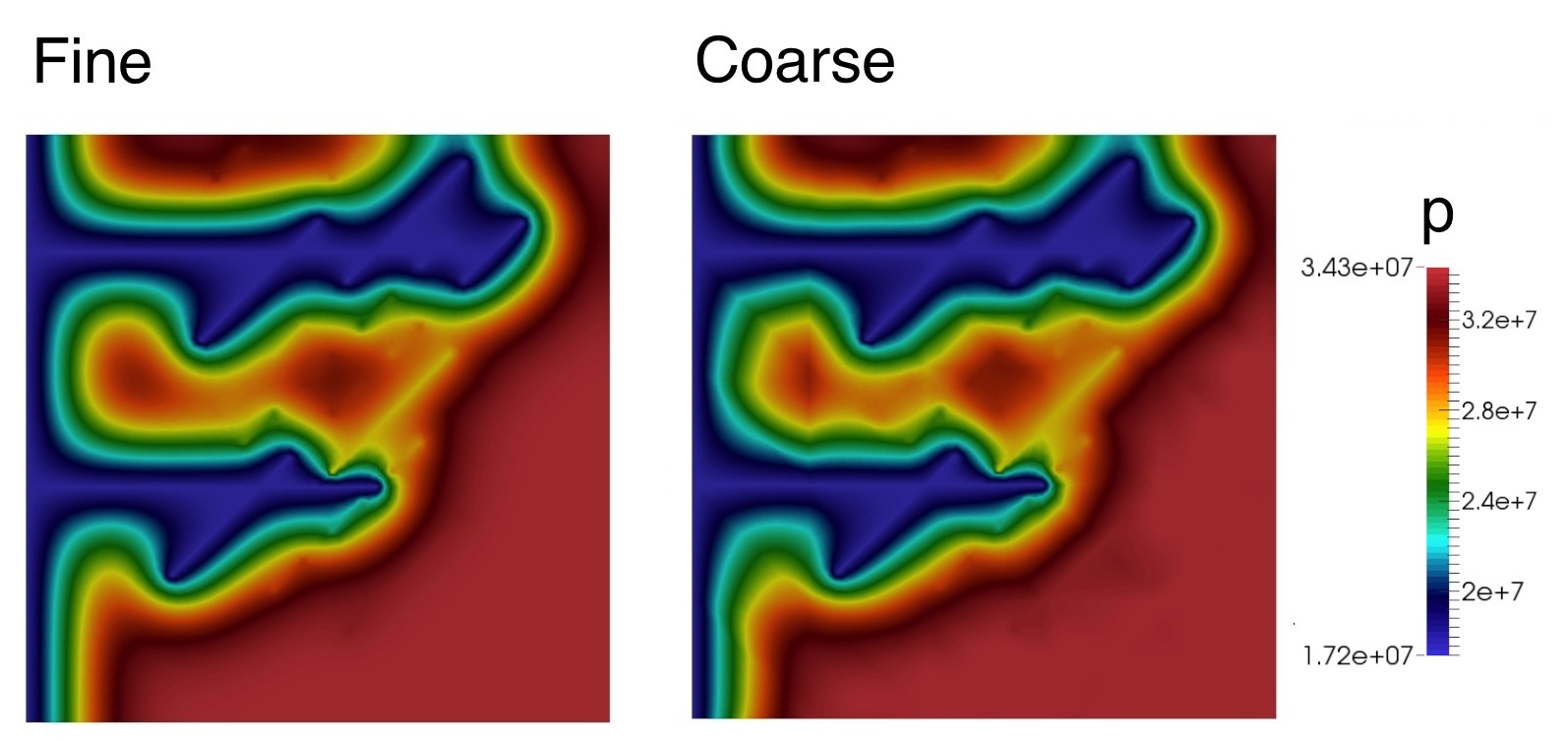}
  \caption{Coarse-scale (top) and fine-scale (bottom) solutions for t=5 year for the case of nonlinear coefficients with $\kappa = \kappa_m$}
  \label{fig:solution-km}
\end{figure}

In Figure \ref{fig:solution-lin}, we show the pressure distribution  for three concrete time level  $t = 1, 3$ and $5$ years. For the pressure and concentration, we have the following relationship: $p = R T c$. Pressure distribution for nonlinear matrix-fracture coefficients in \eqref{eq:discr2} is presented in Figures \ref{fig:solution-k0} - \ref{fig:solution-km} for last time level. In these figures, we show fine-scale (reference) and coarse-scale (multiscale) solutions. 
The coarse-scale solution is obtained in an offline space of dimension $288$ (using $M_{off} = 8$ multiscale basis functions per coarse neighborhood) and the fine-scale solution is obtained in a space of dimension $3891$.
Compared  to the fine-scale solution on the left with the coarse-scale solution on the right of the figures, we observe that the GMsFEM can approximate the fine-scale solution accurately.

To compare the results, we use relative weighted errors
\[
||\varepsilon||_* = ||c_{\text{ms}} - c_h||_* / || c_h ||_*,
\]
using  $L^2_a$ and $H^1_a$ weighted norms that are defined as
\[
||\varepsilon||_{L^2_a} =\left( \int_{\Omega} a \, \varepsilon^2 \, dx \right)^{1/2}, \quad 
||\varepsilon||_{H^1_a} =\left( \int_{\Omega} (a \,  \nabla \varepsilon, \nabla \varepsilon) \, dx \right)^{1/2}.
\]

\begin{figure}[htb]
  \centering
  \includegraphics[width=0.95 \textwidth]{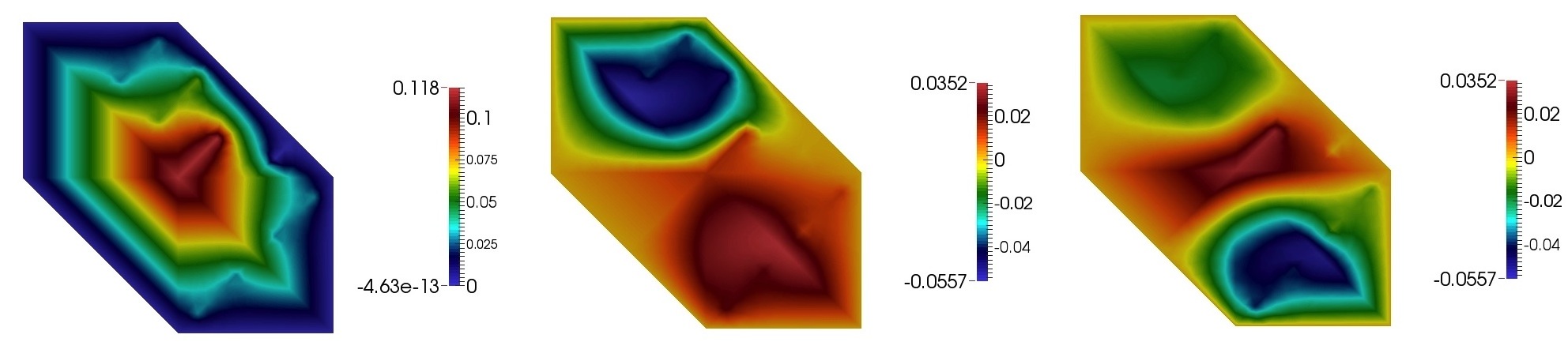}
  \caption{Multiscale basis functions corresponding to the first $3$
smallest eigenvalues in the case with constant fracture-matrix properties after multiplication to partition of unity functions, $\psi_{i,k} = \chi_i \psi_k^{\omega_i, \text{off}}$, $i = 25$ and $k=0,1,2$ (from left to right)}
  \label{fig:eigen}
\end{figure}

In Table \ref{tab:lin-5-10}, we present relative errors (in percentage) for last time level for 
constant fracture and matrix properties in \eqref{eq:discr2} 
using coarse grids with 36 and 121 nodes.
For the coarse-scale approximation, we vary the dimension of the spaces 
by selecting a certain number of offline basis functions ($M_{off}$) corresponding to the smallest eigenvalues. 
In the Table \ref{tab:lin-5-10}, we recall that $V_{off}$ denotes the offline space, $dim(V_{off})$ is the offline space dimension, $M_{off}$ is the number of the multiscale basis functions per coarse neighborhood (we use a similar number of $M_{off}$ for each $\omega_i$), $c_{\text{ms}}$ and $c_h$ are the multiscale and reference solutions, respectively. 

\begin{figure}[htb]
  \centering
  \includegraphics[width=0.45 \textwidth]{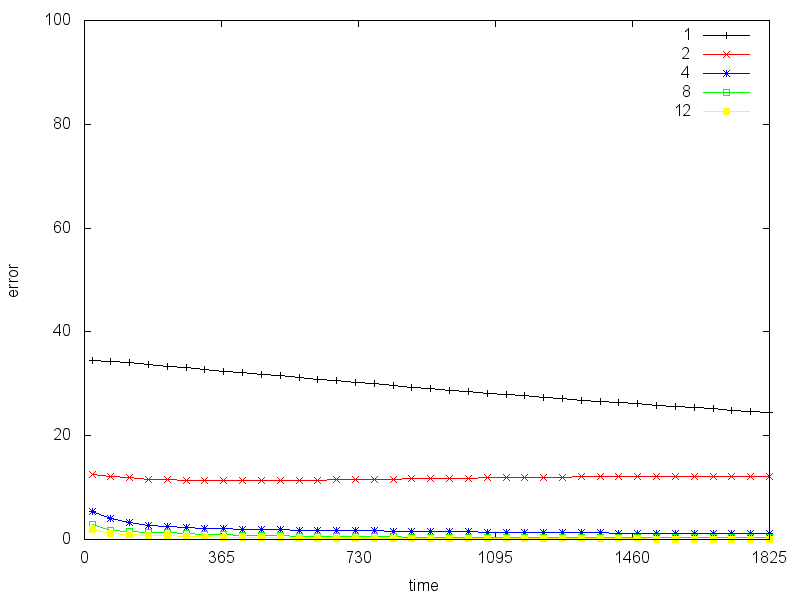}
  \includegraphics[width=0.45 \textwidth]{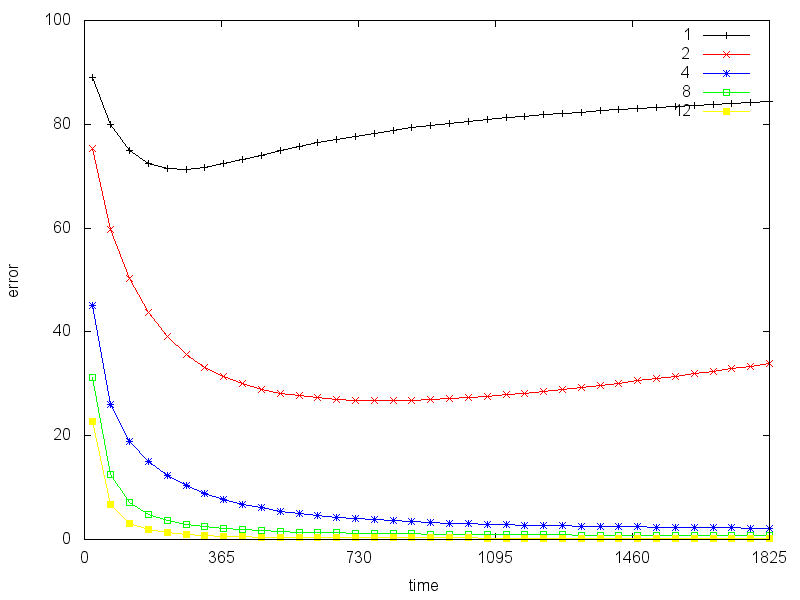}
  \caption{Relative $L^2_a$ and $H^1_a$ weighted errors (\%)  for coarse grid in Figure \ref{fig:meshes5} with 36 nodes. Constant matrix-fracture properties.}
  \label{fig:error5}
\end{figure}

\begin{figure}[htb]
  \centering
  \includegraphics[width=0.45 \textwidth]{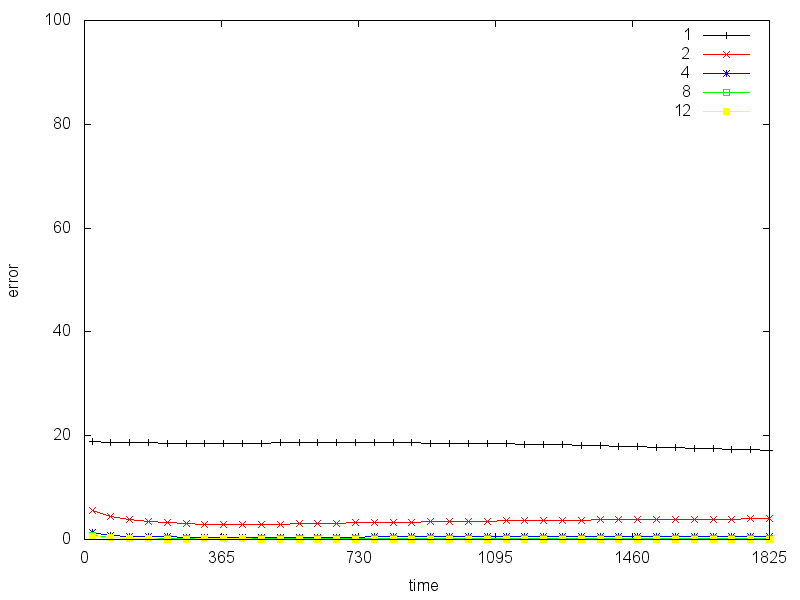}
  \includegraphics[width=0.45 \textwidth]{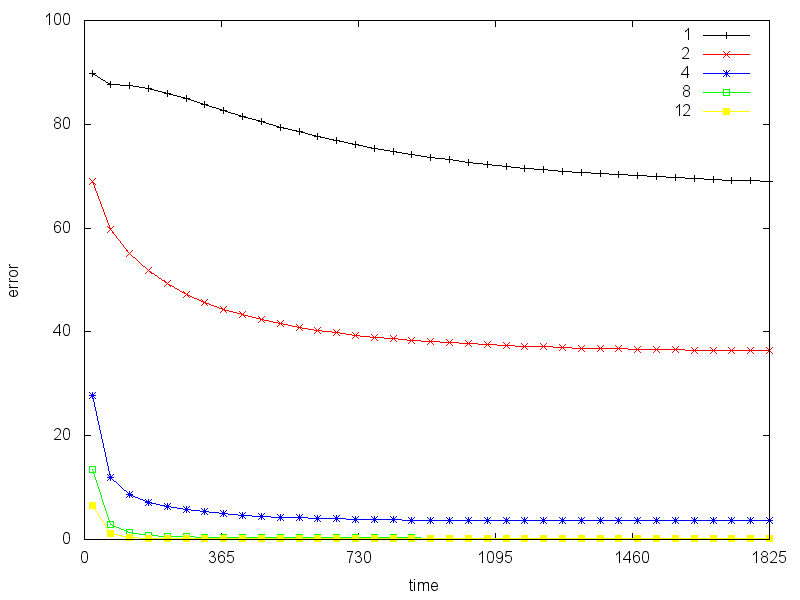}
  \caption{Relative $L^2_a$ and $H^1_a$ weighted errors (\%)  for coarse grid in Figure \ref{fig:meshes5} with 121 nodes. Constant matrix-fracture properties.}
  \label{fig:error10}
\end{figure}

Figure \ref{fig:eigen} presents the multiscale basis functions corresponding to the first $3$ smallest eigenvalues in the case with constant fracture-matrix properties in \eqref{eq:discr2}.
These offline basis functions are multiplied by 
partition of unity functions. 
When we use $M_{off} = 8$ and the case with $36$ coarse nodes, the relative $L^2_a$ and $H^1_a$ weighted errors are $0.3$ \% and  $0.7$\%, respectively for final time level. The dimension of the corresponding offline space is $288$ and for reference solution is $3891$. For coarse grid with $121$ nodes, the relative errors are slightly smaller $0.1$\% and  $0.2$\% for  $L^2_a$ and $H^1_a$ weighted errors, respectively. The dimension of the corresponding offline space is $968$ and for reference solution is $6659$.  The  relative $L^2_a$ and $H^1_a$ errors at different time instants for the cases with $36$ and $121$ coarse grids  are presented in Figures \ref{fig:error5} and \ref{fig:error10}. As we observe
if we take $4$ or more basis functions per coarse node, the relative errors
remain small.
 
\begin{table}[htp]
\begin{center}
\begin{tabular}[hp]{|c|c|c|cc|}
\hline
$M_{off}$  & dim($V_{off}$) & $\lambda_{min}$ & $L^2_a$ & $H^1_a$ \\
\hline \hline
1		& 36		& 9.0 10$^{-9}$		&	24.484  	&	 84.383\\
2     	& 72 	& 4.5 10$^{-8}$		&	12.229 	&	 33.923\\
4     	& 144 	& 1.1 10$^{-7}$		&	1.068 		&	 2.162\\
8     	& 288 	& 2.2 10$^{-6}$		&	0.303 		&	 0.737\\
12   	& 432 	& 0.19						&	0.083 		&	 0.258\\
\hline 
\end{tabular}
$\;\;\;$  
\begin{tabular}[hp]{|c|c|c|c|c|}
\hline
$M_{off}$  & dim($V_{off}$) & $\lambda_{min}$ & $L^2_a$ & $H^1_a$ \\
\hline \hline
1		& 121		& 2.5 10$^{-8}$		&	17.136 &	 68.989\\
2     	& 242 		& 9.6 10$^{-8}$		&	3.975 	&	 36.337\\
4     	& 484 		& 1.6 10$^{-7}$		&	0.651 	&	 3.595\\
8     	& 968 		& 0.37						&	0.110 	&	 0.246\\
12   	& 1452 	& 1.23						&	0.060 	&	 0.108\\
\hline 
\end{tabular}
\end{center}
\caption{Numerical results (relative errors (\%) for the final time level). Left:  for the case with $36$ coarse nodes. Right:  
for the case with $121$ coarse nodes.}
\label{tab:lin-5-10}
\end{table}

We present relative weighted errors  in Tables \ref{tab:error-k0} and \ref{tab:error-km} for different number of eigenvectors $M_{off}$ for the case with nonlinear matrix-fracture coefficients in \eqref{eq:discr2}.
We consider a case with $36$ coarse nodes.
When we use $M_{off} = 8$ and the case with   $\kappa=\kappa_0$, the relative $L^2_a$ and $H^1_a$ errors are $0.2$ \% and  $0.7$\%, respectively. The dimension of the corresponding offline space is $288$ and for reference solution is $3891$. For the case with $\kappa=\kappa_m$ in \eqref{coeff-ab-m}, 
we have $0.4$\% and  $1.0$\% of relative $L^2_a$ and $H^1_a$ errors, respectively. 
The dimension of coarse spaces for the corresponding number of eigenvectors are $72$, $144$, $288$, $432$ and $576$ for 
$M_{off}= 2, 4, 8$ and $12$.
We observe that as the dimension of the coarse space (the number of selected eigenvectors $M_{off}$) increases, the respective relative errors decrease. Also we have similar error behaviour as for case with constant matrix-fracture coefficients. Moreover, we see that the decrease in the relative error is  fast initially and
one can obtain small errors using only a few basis functions.

\begin{table}[htp]
\begin{center}
\begin{tabular}[hp]{|c|c|c|cc|}
\hline
$M_{off}$  & dim($V_{off}$) & $\lambda_{min}$ & $L^2_a$ & $H^1_a$ \\
\hline \hline
1		& 36		& 4.8 10$^{-9}$		&	21.717 &	 87.897\\
2     	& 72 	& 2.4 10$^{-8}$		&	10.772 &	 38.774\\
4     	& 144 	& 6.0 10$^{-8}$		&	0.933 	&	 1.947\\
8     	& 288 	& 1.1 10$^{-6}$		&	0.270 	&	  0.737\\
12   	& 432 	& 0.19						&	0.123 	&	  0.323\\
\hline 
\end{tabular}
$\;\;\;$  
\begin{tabular}[hp]{|c|c|c|c|c|}
\hline
$M_{off}$  & dim($V_{off}$) & $\lambda_{min}$ & $L^2_a$ & $H^1_a$ \\
\hline \hline
1		& 121		& 2.5 10$^{-8}$		&	14.333 &	  64.197\\
2     	& 242 		& 9.6 10$^{-8}$		&	3.673 	&	  30.510\\
4     	& 484 		& 1.6 10$^{-7}$		&	0.646 	&	  3.272\\
8     	& 968 		& 0.37						&	0.110 	&	  0.251\\
12   	& 1452 	& 1.23						&	0.063 	&	  0.159\\
\hline 
\end{tabular}
\end{center}
\caption{Numerical results (relative weighted errors (\%) for final time level)  for case with $\kappa = \kappa_0$  in \eqref{coeff-ab-m}. Left:  the case with $36$ coarse nodes.  Right:  the case with $121$ coarse nodes.}
\label{tab:error-k0}
\end{table}

\begin{table}[htp]
\begin{center}
\begin{tabular}[hp]{|c|c|c|cc|}
\hline
$M_{off}$  & dim($V_{off}$) & $\lambda_{min}$ & $L^2_a$ & $H^1_a$ \\
\hline \hline
1		& 36		& 2.0 10$^{-9}$		&	24.484 &	  92.039\\
2     	& 72 	& 1.0 10$^{-8}$		&	10.785 &	  35.874\\
4     	& 144 	& 2.6 10$^{-8}$		&	1.247 	&	  2.423\\
8     	& 288 	& 5.1 10$^{-7}$		&  0.432 	&	  1.069\\
12   	& 432 	& 0.19						&	0.234 	&	  0.712\\
\hline 
\end{tabular}
$\;\;\;$  
\begin{tabular}[hp]{|c|c|c|c|c|}
\hline
$M_{off}$  & dim($V_{off}$) & $\lambda_{min}$ & $L^2_a$ & $H^1_a$ \\
\hline \hline
1		& 121		& 2.5 10$^{-8}$		&	16.318 &	  60.267\\
2     	& 242 		& 9.6 10$^{-8}$		&	3.715 	&	  22.872\\
4     	& 484 		& 1.6 10$^{-7}$		&	0.645 	&	  3.000\\
8     	& 968 		& 0.37						&	0.134 	&	  0.386\\
12   	& 1452 	& 1.23						&	0.096 	&	  0.282\\
\hline 
\end{tabular}
\end{center}
\caption{Numerical results (relative weighted errors (\%) for the final time level) for case with $\kappa = \kappa_m$  in \eqref{coeff-ab-m}. Left: the case with $36$ coarse nodes. Right: the case with $121$ coarse nodes.}
\label{tab:error-km}
\end{table}

\begin{remark}

In our numerical simulations, we do not use empricial interpolation
procedures for approximating the nonlinear functionals 
$a_{\cdot}(c,\cdot)$ and $b_{\cdot}(c,\cdot)$ (see \cite{calo2014multiscale}
for more details). In the approaches of \cite{calo2014multiscale}, 
 empirical interpolation concepts \cite{ct_POD_DEIM11} are used to evaluate the
nonlinear functions by  dividing the computation of the nonlinear
function into coarse regions, evaluating the contributions of
nonlinear functions in each coarse region taking advantage of a
reduced-order representation of the solution. By using these approaches,
we can reduce the computational cost associated with evaluating
the nonlinear functions and consequently making the computational
cost to be independent of the fine grid.

\end{remark}

\section{Randomized oversampling GMsFEM}

Next, we present numerical results for the oversampling and the randomized snapshots that can substantially
 save the computational cost for snapshot calculations. 
In this algorithm, instead of solving local harmonic  problems  (\ref{harmonic_ex}) for each fine grid node on the boundary, we solve a small number of harmonic extension local problems with  random boundary conditions \cite{randomized2014}. More precisely, we let 
\[
\psi_{j}^{\omega_i, \text{rsnap}}=r_j, \quad x \in \partial \omega_i^+,
\]
where $r_j$ are independent identical distributed standard Gaussian random vectors on the fine grid nodes of the boundary. 
When we use randomized snapshots, we only generate a fraction of the snapshot vectors by using random boundary conditions.

\begin{figure}[htb]
  \centering
  \includegraphics[width=0.45 \textwidth]{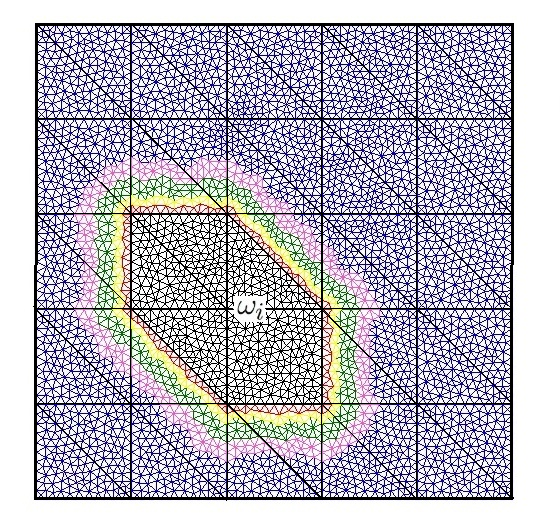}
  \caption{Neighborhood domain with oversampling ($\omega^+_i = \omega_i+m$, $m=1, 2, 4, 6$) for the coarse grid with 36 nodes}
  \label{fig:wplus}
\end{figure}

For snapshot space calculations, we use the extended coarse grid neighborhood for $m=1,2,\dots$, by $\omega^+_i=\omega_{i}+m$, where $m$ is width of the fine-grid layer. Here, for example, $\omega^+_i=\omega_{i}+1$ means the coarse grid neighborhood plus all 1 layer of adjacent fine grid of $\omega_{i}$, and so on (see Figure  \ref{fig:wplus} for illustration). Calculations in the oversampled neighborhood domain $\omega_i^+$ reduces the effects due to the artificial oscillation in random boundary conditions.

\subsection{Numerical results}

The simulation results are presented in Tables \ref{tab:rand5-1} - \ref{tab:rand5-2} for 36 node coarse grid case. We use constant matrix-fracture properties, see  \eqref{eq:discr2}.  We present the results for the randomized snapshot case for last time level.  In our simulations, we set the oversampling size $m=0, 2, 4, 6$ for $\omega^+_i = \omega_i+m$ and use different numbers of multiscale basis functions $M_{off} = 2, 4, 8$ and $12$.

In Table \ref{tab:rand5-1},  we investigate  the effects of the oversampling $\omega^+_i = \omega_i + m$, as we increase the number of fine grid extensions  $m = 0, 2, 4$ and $6$.  We see that  the oversampling  helps to improve the results initially, 
 but the improvements slow and larger oversampling domains do not give significant improvement in the solution accuracy. 
When we use a snapshot ratio of $25.6$ \% (between the standard number of snapshots and the randomized algorithm for $\omega^+_i = \omega_i + 4$), the relative $L^2_a$ and $H^1_a$ weighted errors are $0.2$ \% and  $0.8$\% for full snapshots and  $0.2$ \% and  $0.9$\% for randomized snapshots. We observe that the randomized algorithm can give similar errors as a full snapshots.

 Table \ref{tab:rand5-2} shows relative $L^2_a$ and $H^1_a$ errors for different number of randomized snapshots $M_i$.
The oversampled region $\omega^+_i = \omega_i + 4$  is chosen, that is,  the oversampled region contains an extra $4$ 
fine-grid cell layers around $\omega_i$. 
Our numerical results show that one can achieve a similar accuracy when using a fraction of snapshots with randomized algorithms and thus, it can provide a substantial CPU savings.

\begin{table}
\centering
\begin{tabular}[hp]{|c|cc|cc|}
\hline
\multirow{2}{*}{{$M_{off}$}}  &
\multicolumn{2}{c|}{full snapshots} &
\multicolumn{2}{c|}{randomized snapshots} \\
 & $L^2_a$ & $H^1_a$ & $L^2_a$ & $H^1_a$ \\
 \hline \hline
\multicolumn{5}{|c|}{without oversampling, $\omega_i$} \\ \hline
 &
\multicolumn{2}{c|}{100 \%} &
\multicolumn{2}{c|}{39.7 \%} \\  \hline
2 	 	& 12.229 	&	 33.923 &	 8.303  	&	  33.237\\
4 	 	& 1.068 		&	 2.162 	&	 1.704  	&	  4.730\\
8 	 	& 0.303 		&	 0.737 	&	 1.005  	&	  2.962\\
12 	 & 0.083 		&	 0.258 	&	 0.557  	&	  1.643\\
\hline\hline
\multicolumn{5}{|c|}{with oversampling, $\omega_i^+ = \omega_i+2$} \\ \hline
 &
\multicolumn{2}{c|}{100 \%} &
\multicolumn{2}{c|}{28.3 \%} \\  \hline
2 	 & 12.247 &	 33.943 &	8.921  &	   33.399\\
4 	 & 1.073 	&	 4.237 	&	0.972  &	   3.750\\
8 	 & 0.261 	&	 0.744 	&	0.354  &	   1.003\\
12 & 0.114 	&	 0.329 	&	0.219  &	   0.704\\
\hline\hline
\multicolumn{5}{|c|}{with oversampling, $\omega_i^+ = \omega_i+4$} \\ \hline
 &
\multicolumn{2}{c|}{100 \%} &
\multicolumn{2}{c|}{25.6 \%} \\  \hline
2 	 & 12.216 &	  33.657 	&	 9.334  &	  28.213\\
4 	 & 1.015 	&	  4.576		&	 0.626  &	  2.561\\
8 	 & 0.262 	&	  0.841		&	 0.264  &	  0.949\\
12 & 0.114 	&	  0.349		&	 0.153  &	  0.441\\
\hline\hline
\multicolumn{5}{|c|}{with oversampling, $\omega_i^+ = \omega_i+6$} \\ \hline
 &
\multicolumn{2}{c|}{100 \%} &
\multicolumn{2}{c|}{22.5 \%} \\  \hline
2 	 & 12.746 &	   35.899 	&	  9.455 	&	    27.922\\
4 	 & 1.013 	&	   5.014 	&	  0.603 	&	    2.377\\
8 	 & 0.251 	&	   0.820 	&	  0.277 	&	    0.875\\
12 & 0.124 	&	   0.369 	&	  0.120 	&	    0.421\\
\hline
\end{tabular}
\caption{Randomized oversampling for GMsFEM with number of snapshots  $M_i=24$ (constant matrix-fracture properties) in every $\omega_i^+ = \omega_i+n$, $n=0, 2, 4, 6$ for coarse mesh with 36 nodes (relative errors (\%) for final time level) }
\label{tab:rand5-1}
\end{table}

\begin{table}
\centering
\begin{tabular}[hp]{|c|cc|cc|cc|cc|cc|}
\hline
\multirow{2}{*}{{$M_{off}$}}  &
\multicolumn{2}{c|}{12.8 \% ($M_i=12$)} &
\multicolumn{2}{c|}{17.0 \% ($M_i=16$)} &
\multicolumn{2}{c|}{21.3 \% ($M_i=20$)} &
\multicolumn{2}{c|}{25.6 \% ($M_i=24$)} &
\multicolumn{2}{c|}{29.8 \% ($M_i=28$)} \\  
 & 
 $L^2$ & $H^1$ & 
  $L^2$ & $H^1$ & 
 $L^2$ & $H^1$ & 
 $L^2$ & $H^1$ & 
 $L^2$ & $H^1$ \\
 \hline \hline
2 	 & 8.228 	&    24.878	&	9.449	&	28.895	&	7.346	&	 22.774&	9.334  &	  28.213&	 9.335	&	  26.973\\
4 	 & 1.908 	&    4.208		&	1.381 	&	3.581	&	0.779	&	 2.692	&	0.626  &	  2.561	&	 0.843	&	  4.439\\
8 	 & 0.861 	&    1.777		&	0.589 	&	1.563	&	0.292	&	 1.189	&	0.264  &	  0.949	&	 0.245	&	  0.894\\
12 &  - 		& - 				& 	  0.313	&	0.781	&	0.217	&	 0.581	&	0.153  &	  0.441	&	 0.110	&	  0.393\\
\hline
\end{tabular}
\caption{Randomized oversampling for GMsFEM with different number of snapshots $M_i=12, 16, 20, 24, 28$ in every $\omega_i^+ = \omega_i + 4$ (constant matrix-fracture properties) for coarse mesh with 36 nodes (relative errors (\%) for final time level) }
\label{tab:rand5-2}
\end{table}

\section{Residual based adaptive online GMsFEM}
\label{sec:online}

In this section, we consider the construction of the online basis functions 
that are used in some regions adaptively to reduce the error significantly. We follow
earlier works
\cite{Chung_adaptive14, chung2015residual}, which were done for
linear time-independent problems.
The online basis functions 
are constructed based on a residual
and take into account distant effects. The construction of online
basis functions is motivated by the analysis.
Using the offline computation, 
we construct multiscale basis functions that can 
be used for any input parameters to solve the problem on the coarse grid. 
The fast convergence due to adding online basis functions depends
on the offline space. It is important that the offline space contains
some essential features of the solution space. In our numerical
simulations, we demonstrate that with a sufficient number of offline
basis functions, we can achieve a rapid convergence for the proposed
online procedure.

First, we derive the error indicator for the error $(c^n - c_{\text{ms}}^n)$  for time-dependent problem (\ref{tvarform}) in the energy norm. Furthermore, we use the error indicator to develop an enrichment algorithm. The error indicator gives an estimate of the local error on the coarse grid region $\omega_i$ and we can then add basis functions to improve the solution.

We assume, as before, $V$ is
 the fine-scale finite element space.
To find the fine-scale solution $c^{n+1} \in V$, we solve (as before)
\begin{equation} 
\label{tvarform-f}
m( \frac{ c^{n+1} - c^n }{\tau}, v) +
a(c^{n+1}, v) = (f, v), \quad \forall  v \in V
\end{equation}
and for multiscale solution $c_{\text{ms}}^{n+1} \in V_{\text{off}}$ we have
\begin{equation} 
\label{tvarform}
m( \frac{ c_{\text{ms}}^{n+1} - c_{\text{ms}}^n }{\tau}, v) +
a(c_{\text{ms}}^{n+1}, v) = (f, v), \quad \forall  v \in V_{\text{off}}.
\end{equation}

We define a linear functional $r^n(v)$ for $n$-th time level by
\[
r^n(v) = \tau (f, v) - m(  c_{\text{ms}}^{n+1} - c_{\text{ms}}^n, v) - \tau a(c_{\text{ms}}^{n+1}, v).
\]
Let $\omega_i$ be a coarse region and $V_i = H^1_0(\omega_i)$  then
\[
\begin{split}
r^n_i(v) 
&= \tau \int_{\omega_i} f v 
- \int_{\omega_i^m}   a_{m} \, (c_{\text{ms}}^{n+1} - c_{\text{ms}}^n) \,  v \, dx 
- \tau \int_{\omega_i^m}  b_{m} \, \nabla c_{\text{ms}}^{n+1}  \cdot \nabla v \, dx \\
&
- d_j \sum\limits_j 
\int_{\omega^{f,j}_i} a_{f}(c) \, (c_{\text{ms}}^{n+1} - c_{\text{ms}}^n)  \, v \, dx 
-  \tau \, d_j \sum\limits_j 
\int_{\omega^{f,j}_i}  b_{f} \, \nabla c_{\text{ms}}^{n+1}  \cdot \nabla v \, dx,
\end{split}
\]
where $\omega_i = \omega_i^m \oplus_j \, d_j \omega_i^{f,j}$
and  $m$ and $f$ represent the matrix and the fracture.

The solution at $(n+1)$ time level ($c_{\text{ms}}^{n+1}$) is the solution of the elliptic problem of the form
\begin{equation} 
\label{tproblem}
a_{\tau}(c^{n+1}_{\text{ms}}, v) = \tau (f, v) + m(c^n_{\text{ms}}, v).
\end{equation}
We use  following notation
\[
a_{\tau} (u, v) = m(u, v) + \tau  a(u, v).
\]
Error estimators for the spatial discretization error take into account the dependence of the elliptic problem (\ref{tproblem}) on the time step parameter $\tau$ and we will use the $\tau$-weighted $H_1$ norm 
\[
||v||^2_{\tau} = \tau ||v||_a^2 + ||v||^2_m,
\]
where 
\[
||v||^2_{\tau} = a_{\tau} (v, v), \quad
||v||^2_{a} = a(v, v), \quad 
||v||^2_{m} =m(v, v).
\]

We define the projection $\Pi: V \rightarrow V_{off}$ by 
\[
\Pi v = \sum_{i=0}^N \chi_i (P_i v),
\]
where 
$P_i: V \rightarrow \text{span} \{ \psi_k^{w_i, off} \}$  be the projection defined by
\[
P_i v = \sum_{k=1}^{l_i} \left( b \, v \, \psi_k^{w_i, off}  \right) \psi_k^{w_i, off}.
\]
The projection $P_i$ is the first $l_i$ terms of spectral expansion in terms of eigenfunctions of following problem
\begin{equation}
\label{eq:neweig1}
a_{\tau}^{\omega_i} ( \Psi^{\text{off}}, v) = \lambda^{\text{off}} \tau \int_{\omega_i} b \, |\nabla \chi_i|^2 \, \Psi^{\text{off}}  \, v \, dx.
\end{equation}
Then 
\[
\tau \int_{\omega_i} b \, |\nabla \chi_i|^2 \, (v - P_i v)^2   \, dx 
\leq 
\frac{1}{\lambda_{l_{i+1}}} a_{\tau}^{\omega_i}  (v - P_i v, v - P_i v).
\] 
and
\[
a_{\tau}^{\omega_i}  (v - P_i v, v - P_i v)   \leq a_{\tau}^{\omega_i}  (v, v).
\]
We note that this spectral problem is different from the original one
formulated in  (\ref{offeig}); however, it involves similar terms, such
as energy norms and $L^2$ norms.

Let $e^n = c^n - c_{\text{ms}}^n$ is error for $n$-th time level and using (\ref{tvarform-f}) and (\ref{tvarform}), we have
\begin{equation}
\label{err1}
m( e^{n+1} - e^n, v) + \tau a(e^{n+1}, v) = (r^n, v),
\end{equation}
where the right hand side can be written as follows
\[
\begin{split}
r^n (v) &= \tau (f, v) - m(  c_{\text{ms}}^{n+1} - c_{\text{ms}}^n, v) - \tau a(c_{\text{ms}}^{n+1}, v)   \\
& \leq \tau (f, v - \Pi v) + \tau (f,  \Pi v)  
- m(  c_{\text{ms}}^{n+1} - c_{\text{ms}}^n, \Pi v) - \tau a(c_{\text{ms}}^{n+1}, \Pi v)  \\ 
& - m(  c_{\text{ms}}^{n+1} - c_{\text{ms}}^n, v - \Pi v) - \tau a(c_{\text{ms}}^{n+1}, v - \Pi v)   \\
& =  \tau (f, v - \Pi v) - m(  c_{\text{ms}}^{n+1} - c_{\text{ms}}^n, v - \Pi v) - \tau a(c_{\text{ms}}^{n+1}, v - \Pi v)  \\
& =\sum_{i=1}^N r_i^n \left( \chi_i (v - P_i v) \right) 
\leq \sum_{i=1}^N ||r_i^n||_{*} ||\chi_i (v - P_i v) ||_{\tau}.
\end{split}
\]
We have
\begin{equation}
\label{est1}
\begin{split}
||\chi_i (v - P_i v) ||_{\tau}^2  
& = ||\chi_i (v - P_i v) ||_{m}^2 
+ \tau ||\chi_i (v - P_i v) ||_{a}^2  \\
& = ||\chi_i (v - P_i v) ||_{m}^2 
+ \tau C \int_{\omega_i} b \, |\nabla \chi_i|^2 \, (v - P_i v)^2 \, dx
+ \tau C \int_{\omega_i} b \, \chi_i^2 \, | \nabla (v - P_i v) |^2 \, dx  \\
& \leq C a_{\tau}^{\omega_i} (v - P_i v, v - P_i v) 
+ \tau C \int_{\omega_i} b \, |\nabla \chi_i|^2 \, (v - P_i v)^2 \, dx  \\
&\leq  \left( C + \frac{C}{\lambda_{l_{i+1}}}  \right)
  a_{\tau}^{\omega_i} (v - P_i v, v - P_i v) 
 \leq 
 \left( C + \frac{C}{\lambda_{l_{i+1}}}  \right)
 a_{\tau}^{\omega_i} (v, v).
 \end{split}
\end{equation}
Therefore
\[
\begin{split}
m( e^{n+1} - e^n, v) & + \tau a(e^{n+1}, v)   \leq 
\sum_{i=1}^N ||r_i^n||_* || \chi_i (v - P_i v)||_{\tau} \leq  
\sum_{i=1}^N ||r_i^n||_*  \left( C + \frac{C}{\lambda_{l_{i+1}}} \right)^{1/2}  a_{\tau}^{\omega_i} (v, v)^{1/2}   \\
& \leq
\left( C + \frac{C}{\Lambda_{min}} \right)^{1/2} 
\left( \sum_{i=1}^N ||r_i^n||_*^2 \right)^{1/2}   
\sum_{i=1}^N  a_{\tau}^{\omega_i} (v, v)^{1/2} \leq
\left( C + \frac{C}{\Lambda_{min}} \right)^{1/2} 
\left( \sum_{i=1}^N ||r_i^n||_*^2 \right)^{1/2} 
a_{\tau} (v, v)^{1/2},
\end{split}
\]
where $\Lambda_{min} = \min_i \lambda_{l_{i+1}}$.

Finally, we take $v = e^{n+1}$
\[
\begin{split}
||e^{n+1}||^2_{\tau} & \leq 
\left( C + \frac{C}{\Lambda_{min}} \right)^{1/2} 
\left( \sum_{i=1}^N ||r_i^n||_*^2 \right)^{1/2} 
||e^{n+1}||_{\tau} + m(e^n, e^{n+1})   \\
& \leq \left( C + \frac{C}{\Lambda_{min}} \right)^{1/2} 
\left( \sum_{i=1}^N ||r_i^n||_*^2 \right)^{1/2} 
||e^{n+1}||_{\tau} + ||e^n||_m ||e^{n+1}||_{\tau}.
\end{split}
\]  
Then
\[
||e^{n+1}||_{\tau} \leq \left( C + \frac{C}{\Lambda_{min}} \right)^{1/2} 
\left( \sum_{i=1}^N ||r_i^n||_*^2 \right)^{1/2}  + ||e^n||_m.
\]  
This inequality residuals give a computable indicator of the error $e^{n+1} = c^{n+1} - c_{\text{ms}}^{n+1}$ in the $\tau$-weighted $H_1$ norm.

\begin{remark}

We note that the analysis suggests the use of (\ref{eq:neweig1}) as a local
eigenvalue problem. This eigenvalue problem is ``slightly'' different from
(\ref{offeig}) that we have used earlier. Our numerical simulations show
that the use of (\ref{eq:neweig1}) improves the convergence of
the offline or online procedures slightly in our numerical examples.
 We will use the spectral
problems based on (\ref{offeig}) in our numerical simulations as
it is independent of time stepping.

\end{remark}

Next, we consider online basis construction.
We use the index $m \geq 1$ to represent the enrichment level. At the enrichment level $m$, we use $V^m_{\text{ms}}$ to denote the corresponding space that can contains both offline and online basis functions.  We will consider a strategy for getting the space $V^{m+1}_{\text{ms}}$ from  $V^m_{\text{ms}}$.  By the online basis functions we mean basis functions that are computed during iterative process, contrary 
to offline basis functions that are computed before iterative process. The online basis functions are computed based on some local residuals for the current multiscale solution $c_{\text{ms}}^{n+1, (m)}$.

Let $V^{m+1}_{\text{ms}} = V^{m}_{\text{ms}} + \text{span} \{ \varphi \}$ be the new approximate space that constructed by adding online basis $\varphi \in V_i$ on the $i$-th coarse neighborhood $\omega_i$ and $c_{\text{ms}}^{n+1, (m+1)} \in V^{m+1}_{\text{ms}}$ be the corresponding GMsFEM solution. 

We define $c_{\text{semi}}^{n+1} = c_{\text{semi}}^{n+1} (c_{\text{ms}}^{n, (m)})$, which satisfies 
\[
a_{\tau}(c_{\text{semi}}^{n+1}, v) = \tau (f, v) + m(c^n_{\text{ms}}, v).
\]
For the error $c_{\text{ms}}^{n+1} - c_{\text{semi}}^{n+1}$, we have
\[
\begin{split}
a_{\tau} (c_{\text{ms}}^{n+1} - c_{\text{semi}}^{n+1}, c_{\text{ms}}^{n+1} - c_{\text{semi}}^{n+1}) &= 
a_{\tau} (c_{\text{ms}}^{n+1}, c_{\text{ms}}^{n+1} - c_{\text{semi}}^{n+1}) - a_{\tau} (c_{\text{semi}}^{n+1}, c_{\text{ms}}^{n+1} - c_{\text{semi}}^{n+1}) \\
& = a_{\tau} (c_{\text{ms}}^{n+1}, c_{\text{ms}}^{n+1} - c_{\text{semi}}^{n+1}) - (\tau f, c_{\text{ms}}^{n+1} - c_{\text{semi}}^{n+1}) - m(c_{\text{ms}}^{n}, c_{\text{ms}}^{n+1} - c_{\text{semi}}^{n+1}) \\
& = r(c_{\text{ms}}^{n+1} - c_{\text{semi}}^{n+1}) 
= \sum_i^N r_i( \chi_i ( P_i(c_{\text{semi}}^{n+1} -c_{\text{ms}}^{n+1})
+ c_{\text{ms}}^{n+1} - c_{\text{semi}}^{n+1}) ) \\
& \leq
\sum_i^N ||r_i||_*  ||\chi_i ( P_i(c_{\text{semi}}^{n+1} -c_{\text{ms}}^{n+1})
+ c_{\text{ms}}^{n+1} - c_{\text{semi}}^{n+1})||_{\tau}.
\end{split}
\] 
Using \eqref{est1} we obtain
\begin{equation}
\label{eq:err-semi}
a_{\tau} (c_{\text{ms}}^{n+1} - c_{\text{semi}}^{n+1}, c_{\text{ms}}^{n+1} - c_{\text{semi}}^{n+1}) 
\leq \left( C + \frac{C}{\Lambda_{min}} \right) \sum_i^N ||r_i||_*.
\end{equation}

The solution $c_{\text{ms}}^{n+1, (m+1)}$ satisfies
\[
||c_{\text{ms}}^{n+1, (m+1)} - c_{\text{semi}}^{n+1}||_{\tau}^2 
\leq 
||v-c_{\text{semi}}^{n+1}||_{\tau}^2, \quad 
\forall v \in V_{\text{ms}}^{m+1}.
\]

Taking $v = c_{\text{ms}}^{n+1, (m)} + \alpha \varphi$, we have
\[
||c_{\text{ms}}^{n+1, (m+1)}  - c_{\text{semi}}^{n+1}||_{\tau}^2 
\leq 
||c_{\text{ms}}^{n+1, (m)} + \alpha \varphi  - c_{\text{semi}}^{n+1} ||_{\tau}^2 = 
||c_{\text{ms}}^{n+1, (m)} - c_{\text{semi}}^{n+1}||_{\tau}^2 
+ 2 \alpha \, a_{\tau}^{\omega_i} (c_{\text{ms}}^{n+1, (m)} - c_{\text{semi}}^{n+1}, \varphi) + \alpha^2 a_{\tau}^{\omega_i}  (\varphi, \varphi).
\]
The last two terms in above inequality measure the amount of the reduction in error when the new basis function $\varphi$ is added to the space $V_{\text{ms}}^{m}$. 

For $\alpha = -1$ and $\varphi \in V_i$  is the  solution of
\[
a_{\tau} (\varphi, v) = r(v), \quad \forall v \in V_i.
\] 

Then for $e_{\text{semi}}^{n+1, (m+1)} = c_{\text{ms}}^{n+1, (m+1)}  - c_{\text{semi}}^{n+1}$ we have
\[
\begin{split}
||e_{\text{semi}}^{n+1, (m+1)} ||_{\tau}^2 
& \leq 
||e_{\text{semi}}^{n+1, (m)} ||_{\tau}^2 
- 2 \, a_{\tau}^{\omega_i} (c_{\text{ms}}^{n+1, (m)} - c_{\text{semi}}^{n+1} , \varphi) 
+ r_i (\varphi) \\
& \leq ||e_{\text{semi}}^{n+1, (m)} ||_{\tau}^2 
- 2 \, a_{\tau}^{\omega_i} (c_{\text{ms}}^{n+1, (m)}, \varphi) + 2 \tau (f, v) + 2 m(c^{n, (m)}_{\text{ms}}, \varphi) 
+ r (\varphi) \leq
||e_{\text{semi}}^{n+1, (m)} ||_{\tau}^2 - ||r ||_*^2. 
\end{split}
\]
To enhance the convergence and efficiency of the online adaptive GMsFEM, we consider enrichment on non-overlapping coarse neighbothoods. Let $I \subset \{  1, 2, \dots ,N \}$ be the index set of some non-overlapping coarse neighborhoods. We define $V^{m+1}_{\text{ms}} = V^{m}_{\text{ms}} + \text{span} \{ \varphi_i, \, \i \in I \}$ and obtain
 \[
||e_{\text{semi}}^{n+1, (m+1)} ||_{\tau}^2 
\leq 
||e_{\text{semi}}^{n+1, (m)} ||_{\tau}^2 - \sum_{i \in I} ||r_i ||_*^2.
\]

Finally, we combine this with \eqref{eq:err-semi} and obtain
\[
\begin{split}
||e_{\text{semi}}^{n+1, (m+1)} ||_{\tau}^2 
&\leq 
||e_{\text{semi}}^{n+1, (m)} ||_{\tau}^2 - 
\sum_{i \in I}  || r_i ||^2_{*} 
\frac{a_{\tau}^{\omega_i} (e_{\text{semi}}^{n+1, (m)} , e_{\text{semi}}^{n+1, (m)} )}{ \left( C + \frac{C}{\Lambda_{min}} \right) \sum_{i=1}^N ||r_i||_*^2 } \\
& \leq 
\left( 1 -  \frac{ \sum_{i \in I}  || r_i ||^2_{*}  }{ \left( C + \frac{C}{\Lambda_{min}} \right) \sum_{i=1}^N ||r_i||_*^2 } \right) 
||e_{\text{semi}}^{n+1, (m)} ||_{\tau}^2.
\end{split}
\]

We will find  online basis functions $\varphi \in V_i$ to maximize the local resudial  $r^n_i$ for current time level. Moreover, the required $\varphi$ is the solution of
\begin{equation}
\label{online}
a_{\tau}(\varphi, v) = r_i(v), \quad \forall v \in V_i,
\end{equation}
where $R_i(v)$is the local residual  that defined using $w_{\text{ms}}^{m}$ 
\[
r_i(v) =\tau (f, v) - m(  c_{\text{ms}}^{n+1} - c_{\text{ms}}^n, v) - \tau a(c_{\text{ms}}^{n+1}, v)
\]
and $||r_i||^2_{*} = ||\varphi||^2_{\tau}$ according to the Riez representation theorem.

For solution in each time level, we iteratively enrich our offline space by residual based online basis function. These basis functions are calculated using Equation (\ref{online}) with zero Dirichlet boundary conditions and the residual norm $||r^n_i||_{*}$ provides a measure on the amount of reduction in energy error.

For the construction of the adaptive online basis functions, we first choose $0<\theta<1$, for each coarse neighborhood $\omega_i$, find the online basis $\varphi_i \in V_i$ using equation \eqref{online}.
After compute the norm of local residuals and calculate $\eta_i$
\[
\eta^2_i:= \lVert r_i\rVert _{\star}^2,
\]
where $\lVert r_i\rVert _{\star} =\lVert \phi_i \rVert _{\tau}$, then arrange them in descending order, i.e. $\eta^2_1 \geq \eta^2_2 \geq ... \geq \eta^2_N$.
Then, choose the smallest $k$ such that 
\[
\theta \sum_{i=1}^N \eta^2_i \leq \sum_{i=1}^k \eta^2_i.
\]
This implies that, for the coarse neighborhood $\omega_j (j=1,...k)$, we  add the corresponding online basis $\varphi_j$ to the original space $V_{ms}^m$.

\subsection{Numerical results}

Next, we present numerical results for residual based online basis functions. We consider a similar problem as in the previous section with  constant matrix-fracture properties in \eqref{eq:discr2} and iteratively enrich the offline space by online residual basis functions in some selected time steps. 
Our coarse and fine grid setups are the same as in Section \ref{sec:numerical}.
Because re-generation of the matrix $R$ is needed, when we add online basis function, we add them for some selected time steps. 
We note that, when we add new online basis functions, which are based on current residuals, we remove previously calculated online basis function and keep them till we update the online basis functions. It will save computational time if we have small size of coarse scale problem.

In Table \ref{tab:erronline}, we present $L^2_a$ and $H^1_a$ errors.
We consider three different cases. In the first case (we call it {\bf Case 1}), online basis functions
are added at the first time step and after that in every $30$-th time step.
In the second case (we call it {\bf Case 2}), online basis functions
are replaced at the first five consecutive time steps, and after 
that, the online basis functions are updated  in every $30$-th time step.
In the third case (we call it {\bf Case 3}), online basis functions
are replaced at the first ten consecutive time steps, and after 
that, the online basis functions are updated  in every $30$-th time step.
More updates initially helps to reduce the error due to the initial condition.
As we mentioned that the offline space is important for the convergence, and
we present the results for different number of initial offline basis functions per coarse neighborhood. We use multiscale basis functions from offline space as a initial basis functions. 
In Table \ref{tab:erronline2}, we show errors when 
 online basis functions
are replaced at the first five consecutive time steps (as in Case 2), and
afterwards, online basis functions are updated at 
$10$-th, $20$-th and $30$-th time step. 
For our calculations, we use $t_{max} = 5$ years with $\tau = 10$ days. 
Calculations are performed in the coarse grid with 121 nodes 
for the case with constant matrix-fracture properties.
We observe from this table the following facts.

\begin{itemize}

\item Choosing $4$ initial offline basis functions improves the convergence
substantially. This indicates that the choice of the initial
offline space is important.

\item Adding online basis functions less frequently (such as at every $30$th
time step) provides an accurate approximation of the solution. This indicates
that the online basis functions can be added only at some selected time
steps.

\end{itemize}

\begin{figure}[htb]
  \centering
  \includegraphics[width=0.35 \textwidth]{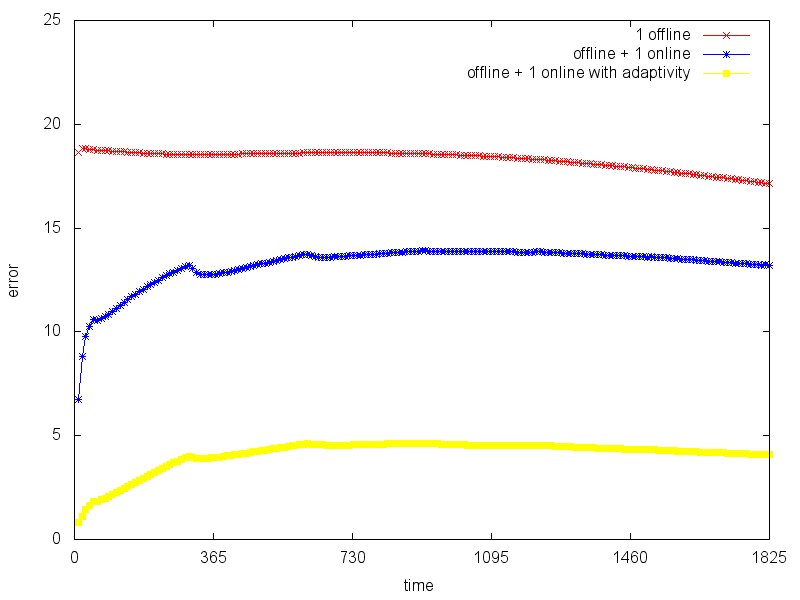}
  \includegraphics[width=0.35 \textwidth]{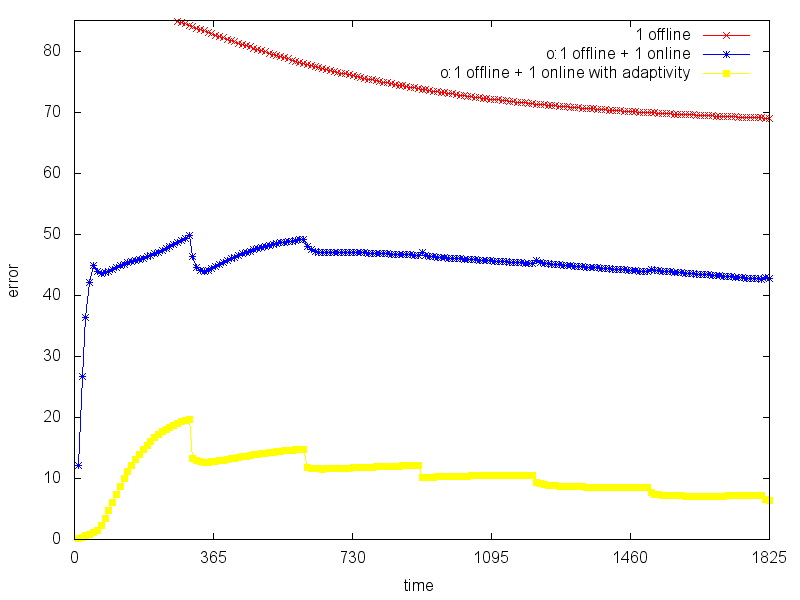}
  \\
  \includegraphics[width=0.35 \textwidth]{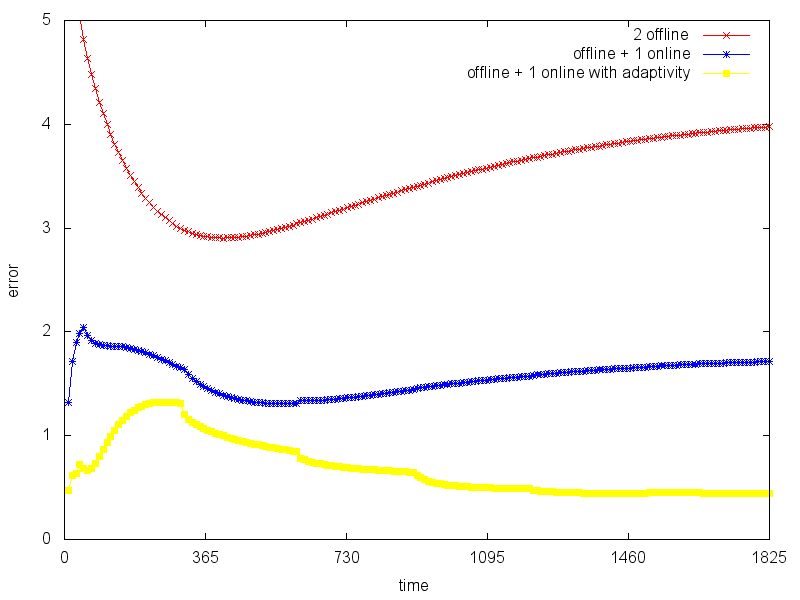}
  \includegraphics[width=0.35 \textwidth]{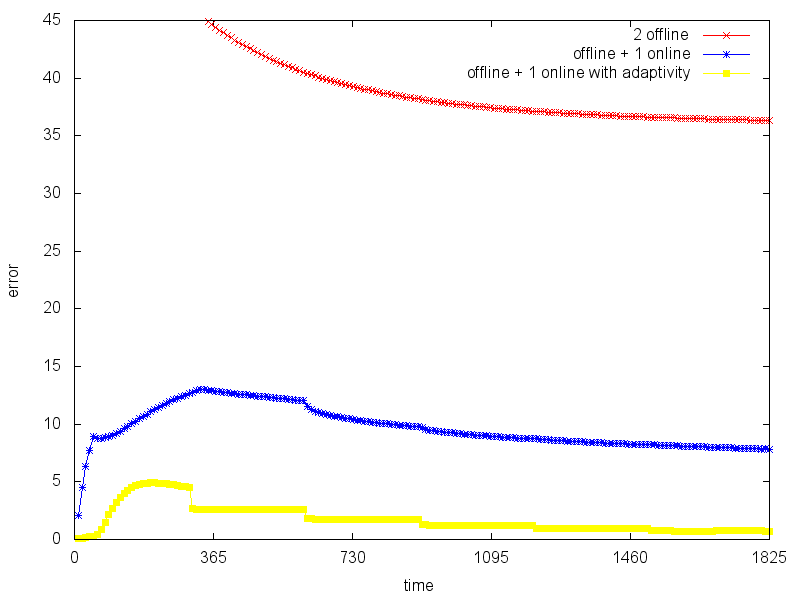}
  \\ 
  \includegraphics[width=0.35 \textwidth]{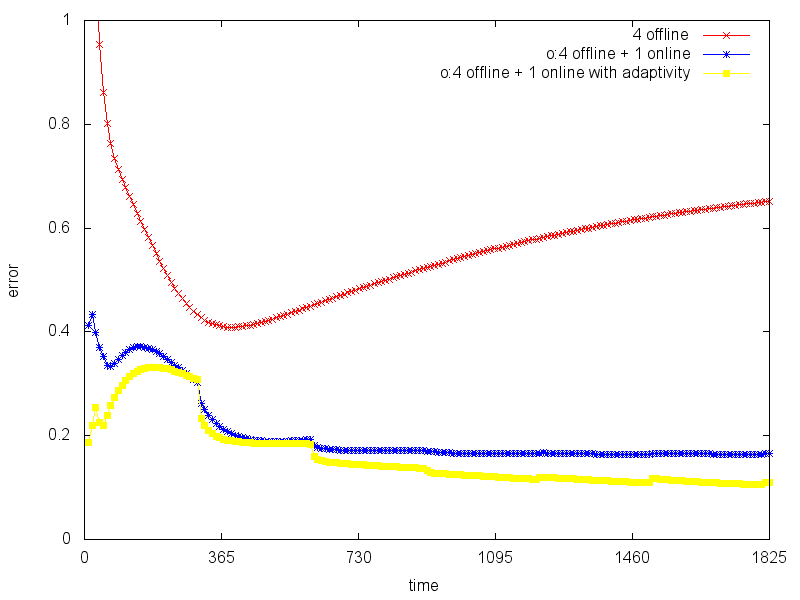}
  \includegraphics[width=0.35 \textwidth]{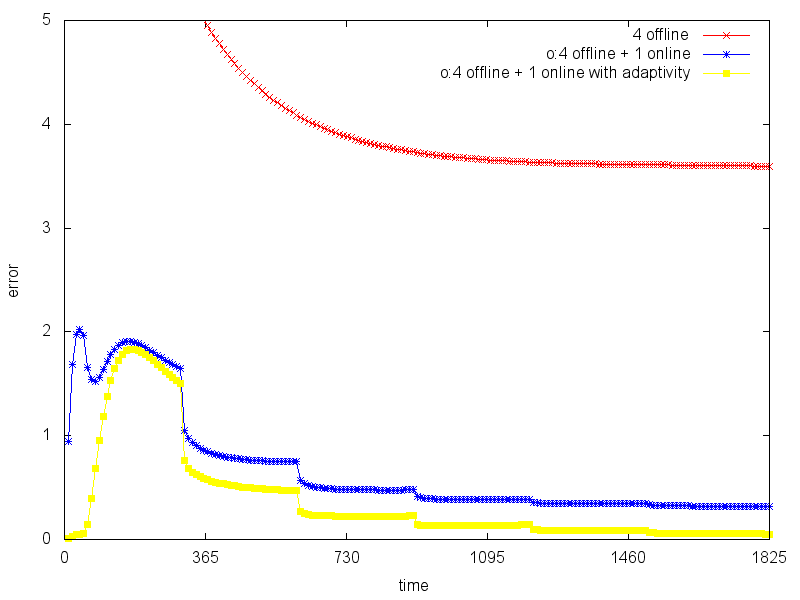}
  \caption{Dynamic of relative $L^2_a$ (left) and $H^1_a$ (right) weighted errors (\%)  for coarse grid in figure \ref{fig:meshes10} with 121 nodes for the case with constant coefficients. Weighted errors using offline basis functions and online basis functions with and without adaptivity. Top: 1 offline basis function. Middle: 2 offline basis functions. Bottom: 4 offline basis functions.}
  \label{fig:error-adaptive}
\end{figure}

Next, we would like to show that one can use online basis functions adaptively
and use the adaptivity criteria discussed above. 
In Table \ref{tab:erronline-space-adaptive},  we present results for residual based online basis functions with adaptivity with $\theta = 0.7$. In Figure  \ref{fig:error-adaptive}, we show errors by time. We observe that applying adaptive algorithm can much reduce errors.

\begin{figure}[htb]
  \centering
  \includegraphics[width=1 \textwidth]{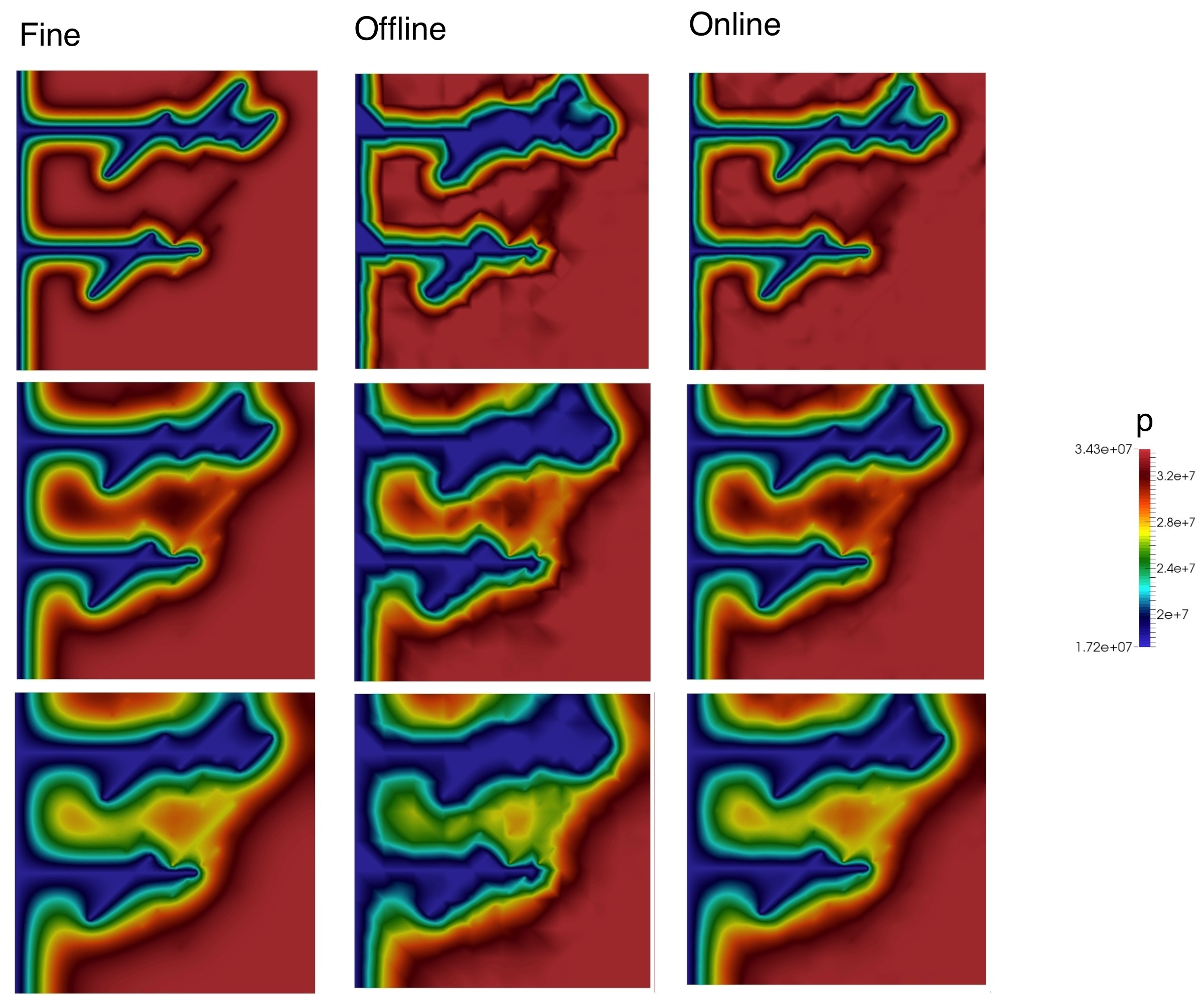}
  \caption{Fine scale solution (right), coarse-scale using 2 offline basis functions (middle) and coarse-scale after two online iteration for some time levels (left) for t=1, 3 and 5 year (from top to bottom) (constant matrix-fracture coefficients). For fine-scale solution size of problem is 6659. For 2 offline basis functions is 242 and after two online iteration is 484}
  \label{fig:solution-lin-on}
\end{figure}

\begin{table}[!htb]
\centering
\begin{tabular}{ |c | c | c | }
\hline
$DOF$ & \multirow{2}{*}{$L^2_a$}  & \multirow{2}{*}{$H^1_a$}  \\
(\# iter)  &	& \\  \hline 
\multicolumn{3}{|c|}{$M_{off} = 1$} \\
\hline
				121 			& 17.136 &	 68.989 \\  \hline
				242 (1) 	& 13.047 &	  43.662 \\  \hline
				363 (2) 	& 7.275 	&	  12.262 \\  \hline
\multicolumn{3}{|c|}{$M_{off} = 2$} \\
\hline
				242 			&	3.975 &	 36.337\\ \hline
				363 (1) 	&	1.653 &	  6.705 \\  \hline
				484 (2) 	&	0.889 &	  0.972 \\  \hline
\multicolumn{3}{|c|}{$M_{off} = 4$} \\
\hline
				484 			&  0.651 &  3.595 \\ \hline
				605 (1) 	&	0.208 &	 0.307 \\  \hline
				726 (2) 	&	0.171 &	  0.056 \\  \hline
\end{tabular} 
$\;\;\;$  
\begin{tabular}{ |c | c | c | }
\hline
$DOF$ & \multirow{2}{*}{$L^2_a$}  & \multirow{2}{*}{$H^1_a$}  \\
(\# iter)  &	& \\  \hline 
\multicolumn{3}{|c|}{$M_{off} = 1$} \\
\hline
				121 			& 17.136 &	 68.989 \\  \hline
				242 (1) 	& 13.209 &	  42.777 \\  \hline
				363 (2) 	& 6.603 	&	  12.689 \\  \hline
\multicolumn{3}{|c|}{$M_{off} = 2$} \\
\hline
				242 			&	3.975 	&	 36.337\\ \hline
				363 (1) 	&	1.716 	&	  7.841 \\  \hline
				484 (2) 	&	0.546 	&	  0.914 \\  \hline
\multicolumn{3}{|c|}{$M_{off} = 4$} \\
\hline
				484 			&  0.651 	&  3.595 \\ \hline
				605 (1) 	&	0.165 	&	  0.313 \\  \hline
				726 (2) 	&	0.105 	&	  0.057 \\  \hline
\end{tabular}
$\;\;\;$  
\begin{tabular}{ |c | c | c | }
\hline
$DOF$ & \multirow{2}{*}{$L^2_a$}  & \multirow{2}{*}{$H^1_a$}  \\
(\# iter)  &	& \\  \hline 
\multicolumn{3}{|c|}{$M_{off} = 1$} \\
\hline
				121 			& 17.136 &	 68.989 \\  \hline
				242 (1) 	& 13.002 	&	  42.091 \\  \hline
				363 (2) 	& 6.125 	&	  13.186 \\  \hline
\multicolumn{3}{|c|}{$M_{off} = 2$} \\
\hline
				242 			&	3.975 	&	 36.337\\ \hline
				363 (1) 	&	1.692 	&	  8.709 \\  \hline
				484 (2) 	&	0.449 	&	  0.862 \\  \hline
\multicolumn{3}{|c|}{$M_{off} = 4$} \\
\hline
				484 			&  0.651 	&  3.595 \\ \hline
				605 (1) 	&	0.144 	&	  0.318 \\  \hline
				726 (2) 	&	0.076 	&	  0.056 \\  \hline
\end{tabular} 
\caption{Convergence history using one, two and four offline basis functions ($M_{off} = 1, 2$ and 4). We add online basis functions for every 30 time step and for  $N-$th first steps (Cases 1, 2, and 3). Left: $N = 1$. Middle: $N = 5$. Right: $N = 10$. Here $DOF$ for the last time step.}
\label{tab:erronline}
\end{table} 

\begin{table}[!htb]
\centering
\begin{tabular}{ |c | c | c | }
\hline
$DOF$ & \multirow{2}{*}{$L^2_a$}  & \multirow{2}{*}{$H^1_a$}  \\
(\# iter)  &	& \\  \hline 
\multicolumn{3}{|c|}{$M_{off} = 1$} \\
\hline
				121 			& 17.136 &	 68.989 \\  \hline
				242 (1) 	& 12.511 &	  41.477 \\  \hline
				363 (2) 	& 5.910 	&	  13.153 \\  \hline
\multicolumn{3}{|c|}{$M_{off} = 2$} \\
\hline
				242 			&	3.975 &	 36.337\\ \hline
				363 (1) 	&	1.624 &	  8.225 \\  \hline
				484 (2) 	& 	0.378 &	  0.934 \\  \hline
\multicolumn{3}{|c|}{$M_{off} = 4$} \\
\hline
				484 			&  0.651 	&  3.595 \\ \hline
				605 (1) 	&	0.126 &	  0.303 \\  \hline
				726 (2) 	&	0.048 &	  0.033 \\  \hline
\end{tabular} 
$\;\;\;$  
\begin{tabular}{ |c | c | c | }
\hline
$DOF$ & \multirow{2}{*}{$L^2_a$}  & \multirow{2}{*}{$H^1_a$}  \\
(\# iter)  &	& \\  \hline 
\multicolumn{3}{|c|}{$M_{off} = 1$} \\
\hline
				121 			& 17.136 &	 68.989 \\  \hline
				242 (1) 	& 12.925	&	  42.778\\ \hline
				363 (2) 	& 6.275	&	   12.705\\ \hline
\multicolumn{3}{|c|}{$M_{off} = 2$} \\
\hline
				242 			&	3.975 	&	 36.337\\ \hline
				363 (1) 	&	1.669	&	   8.034\\ \hline
				484 (2) 	&	0.474	&	   0.880\\ \hline
\multicolumn{3}{|c|}{$M_{off} = 4$} \\
\hline
				484 			&  0.651 	&  3.595 \\ \hline
				605 (1) 	&	0.147	&	   0.305\\ \hline
				726 (2) 	&	0.080	&	   0.041\\ \hline
\end{tabular}
$\;\;\;$  
\begin{tabular}{ |c | c | c | }
\hline
$DOF$ & \multirow{2}{*}{$L^2_a$}  & \multirow{2}{*}{$H^1_a$}  \\
(\# iter)  &	& \\  \hline 
\multicolumn{3}{|c|}{$M_{off} = 1$} \\
\hline
				121 			& 17.136 &	 68.989 \\  \hline
				242 (1) 	& 13.209 &	  42.777 \\  \hline
				363 (2) 	& 6.603 	&	  12.689 \\  \hline
\multicolumn{3}{|c|}{$M_{off} = 2$} \\
\hline
				242 			&	3.975 	&	 36.337\\ \hline
				363 (1) 	&	1.716 	&	  7.841 \\  \hline
				484 (2) 	&	0.546 	&	  0.914 \\  \hline
\multicolumn{3}{|c|}{$M_{off} = 4$} \\
\hline
				484 			&  0.651 	&  3.595 \\ \hline
				605 (1) 	&	0.165 	&	  0.313 \\  \hline
				726 (2) 	&	0.105 	&	  0.057 \\  \hline
\end{tabular}
\caption{Convergence history using one, two and four offline basis functions ($M_{off} = 1, 2$ and $4$). We add online basis functions for every $N$th time step and for first $5$ steps Left: $N = 10$. Middle: $N = 20$. Right: $N = 30$. Here $DOF$ for last time step}
\label{tab:erronline2}
\end{table}

\begin{table}[!htb]
\centering
\begin{tabular}{ |c | c | c | c | }
\hline
$DOF$  &  \multirow{2}{*}{$\sum$ iter} & \multirow{2}{*}{$L^2_a$}  & \multirow{2}{*}{$H^1_a$}  \\
(\# iter)  & 	& & \\  \hline 
\multicolumn{4}{|c|}{$M_{off} = 1$} \\
\hline
				121 			& 			& 17.136 &	 68.989 \\  \hline
				242 (1) 	& 11 	& 13.209 &	  42.777 \\  \hline
				363 (2) 	& 22 	& 6.603 	&	  12.689 \\  \hline
\multicolumn{4}{|c|}{$M_{off} = 2$} \\
\hline
				242 			&			& 3.975 	&	 36.337\\ \hline
				363 (1) 	&	11	& 1.716 	&	  7.841 \\  \hline
				484 (2) 	&	22	& 0.546 	&	  0.914 \\  \hline
\multicolumn{4}{|c|}{$M_{off} = 4$} \\
\hline
				484 			&  		& 0.651 	&  3.595 \\ \hline
				605 (1) 	&	11	& 0.165 	&	  0.313 \\  \hline
				726 (2) 	&	22	& 0.105 	&	  0.057 \\  \hline
\end{tabular}
$\;\;\;$  
\begin{tabular}{ |c | c | c | c | }
\hline
\multirow{2}{*}{$DOF$}  & \multirow{2}{*}{$\sum$ iter}  & \multirow{2}{*}{$L^2_a$}  & \multirow{2}{*}{$H^1_a$}  \\
  & 	& & \\  \hline 
\multicolumn{4}{|c|}{$M_{off} = 1$} \\
\hline
				121 			& 			& 17.136 &	 68.989 \\  \hline
				243  	& 44 	& 4.082 	&	  6.418 \\ \hline
				381  	& 78 	& 2.681 	&	  2.871 \\ \hline
\multicolumn{4}{|c|}{$M_{off} = 2$} \\
\hline
				242 			&			& 3.975 	&	 36.337\\ \hline
				376  	&	44	& 0.441 	&	  0.720 \\ \hline
				504  	&	77	& 0.376 	&	  0.325 \\ \hline
\multicolumn{4}{|c|}{$M_{off} = 4$} \\
\hline
				484 			&  		& 0.651 	&  3.595 \\ \hline
				635 	&	44	& 0.110 	&	  0.044 \\ \hline
				737 	&	68	& 0.098 	&	  0.039 \\ \hline
\end{tabular}
\caption{Convergence history using one, two and four offline basis functions ($M_{off} = 1, 2$ and 4). We add online basis functions for every 30 time step and for first 5 steps Left: without space adaptivity. Right: with space adaptivity. Here $DOF$ for last time step}
\label{tab:erronline-space-adaptive}
\end{table}

\section{Conclusions}

In this paper, we present a multiscale approach for shale transport in fractured media. Our approach uses an upscaled model in
the form of nonlinear parabolic equations to
represent the matrix that consists of organic and inorganic matter. The nonlinearities in the equation are due
to the interaction of organic and inorganic matter. The interaction of nonlinear matrix and the fracture
is represented by multiscale basis functions. We follow Generalized Multiscale Finite Element Method to
extract the leading order terms that represent the matrix and the fracture interaction. Multiscale basis functions
are constructed locally in each coarse region and they represent the interaction between the upscaled matrix
and the fracture network. We show that our proposed approach can effectively capture the small-scale effects
and the overall system can be modeled using a fewer degrees of freedom. Numerical results are presented. In some cases and some regions, 
 the offline procedure is
insufficient to give accurate representations of the solution, due to
the fact that offline computations are typically performed locally and
global information is missing in these offline information.  These
phenomena occur locally and in some of these regions that are
identified using the proposed error indicators, we need to
develop online basis functions~\cite{ chung2015residual}. 
 We  discuss online basis functions and show that this procedure converges
fast.

\section{Acknowledgements}

We are grateful to Tat Leung for helpful discussions and suggestions
regarding to online basis constructions.
MV's work is partially supported by Russian Science Foundation Grant RS 15-11-10024.

\bibliographystyle{plain}
\bibliography{references}
\end{document}